\documentclass[aps,prl,reprint, amsmath,amssymb,
 aps, superscriptaddress, showpacs]{revtex4-1}
\usepackage{graphicx}
\usepackage{braket}

\usepackage{tikz}
\usepackage{pgfplots}
\pgfplotsset{compat=1.8}
\usetikzlibrary{pgfplots.groupplots}
\usepgfplotslibrary{external} 
\pgfplotsset{compat=newest}
\tikzsetexternalprefix{bilder/}
\graphicspath{{bilder/}}
\usepackage{subfigure}
\begin{document}

% Use the \preprint command to place your local institutional report
% number in the upper righthand corner of the title page in preprint mode.
% Multiple \preprint commands are allowed.
% Use the 'preprintnumbers' class option to override journal defaults
% to display numbers if necessary
%\preprint{}

%Title of paper
\title{Fano Resonances in Majorana Bound States  - Quantum Dot Hybrid Systems}

% repeat the \author .. \affiliation  etc. as needed
% \email, \thanks, \homepage, \altaffiliation all apply to the current
% author. Explanatory text should go in the []'s, actual e-mail
% address or url should go in the {}'s for \email and \homepage.
% Please use the appropriate macro foreach each type of information

% \affiliation command applies to all authors since the last
% \affiliation command. The \affiliation command should follow the
% other information
% \affiliation can be followed by \email, \homepage, \thanks as well.
\author{Alexander Schuray}
    \affiliation{Institut f\"ur Mathematische Physik, Technische Universit\"at
        Braunschweig, D-38106 Braunschweig, Germany}
    \author{Luzie Weithofer}
    \affiliation{Institut f\"ur Mathematische Physik, Technische Universit\"at
        Braunschweig, D-38106 Braunschweig, Germany}
    \author{Patrik Recher}
    \affiliation{Institut f\"ur Mathematische Physik, Technische Universit\"at
        Braunschweig, D-38106 Braunschweig, Germany}
    \affiliation{Laboratory for Emerging Nanometrology Braunschweig, D-38106 
        Braunschweig, Germany}
%\email[]{Your e-mail address}
%\homepage[]{Your web page}
%\thanks{}
%\altaffiliation{}
%\affiliation{Institute for Mathematical Physics, TU Braunschweig, 38106 Braunschweig, Germany}

%Collaboration name if desired (requires use of superscriptaddress
%option in \documentclass). \noaffiliation is required (may also be
%used with the \author command).
%\collaboration can be followed by \email, \homepage, \thanks as well.
%\collaboration{}
%\noaffiliation

\date{\today}

\begin{abstract}
We consider a quantum wire, containing two Majorana bound states (MBS) at its ends that are coupled to a current lead on one side and to a quantum dot (QD) on the other side. 
Using the method of full counting statistics we calculate the conductance and the zero-frequency noise. Using an effective low-energy model, we analyze in detail the Andreev reflection probability as a function of the various system parameters and show that it exhibits a Fano 
resonance (FR) line shape in the case of a weakly coupled QD as a function of the QD energy level when the two MBS overlap. The asymmetry parameter changes sign as the bias voltage is tuned through 
the MBS overlap energy. The FR is mirrored as a function of the QD level energy as long as tunneling to the more distant MBS is negligible. However, if both MBS are coupled 
to the lead and the QD, the height as well as the asymmetry of the line shapes cease to respect this symmetry. These two exclusive cases uniquely distinguish the coupling
to a MBS from the coupling to a fermionic bound state that is shared between the two MBS. We complement the analysis by employing a discretized one-dimensional $p$-wave superconductor (Kitaev chain) for the quantum wire and show that the features of the effective low-energy model are robust towards a more complete Hamiltonian and also persist at finite temperature.
\end{abstract}

% insert suggested PACS numbers in braces on next line
% insert suggested keywords - APS authors don't need to do this
%\keywords{}

%\maketitle must follow title, authors, abstract, \pacs, and \keywords
\maketitle

% body of paper here - Use proper section commands
% References should be done using the \cite, \ref, and \label commands
\section{I. Introduction}
Transport through Majorana bound states and their manipulation currently attracts a lot of attention both theoretically and experimentally. These particles,
first proposed in high energy physics as elementary particles being their own antiparticles by Ettore Majorana \cite{Majorana1937}, could represent the basic building
blocks for a topological quantum computer \cite{Nayak2008, Pachos2012}. First experiments have tested their particle-hole symmetry and charge neutrality via a resonance appearing at zero 
energy \cite{Mourik2012, Das2012, Lee2013, Nadj-Perge2014,Pawlak2015}. Another set of experiments investigate the predicted fractional Josephson effect \cite{Kitaev2001,Fu2009b,PhysRevLett.110.017003,PhysRevLett.112.077002} through
the missing odd steps in a Shapiro staircase \cite{Rokhinson2012,Wiedenmann2016}.
An important next step is to perform the braiding operations and to 
show their non-abelian nature \cite{Nayak2008, DasSarma2005, Stern2006, Bonderson2006, Alicea2011, Sau2011, vanHeck2012, Cheng2012, Pedrocchi2015}. The so far best studied systems containing Majorana bound states (MBS) are semiconducting quantum wires \cite{Lutchyn2010,Oreg2010,PhysRevLett.107.196804,PhysRevB.87.024515} having, in the
most ideal case, two MBS with one MBS at each end of the wire. In principle, there are several ways in which a splitting of two MBS can be generated. Either direct wave function overlap or charging effects
can lead to striking transport features like dominant crossed Andreev reflection \cite{Nilsson2008} or teleportation of charge \cite{Fu2010, Albrecht2016}. In addition, a dynamical splitting can be induced by the braiding operation of two MBS in a Corbino geometry topological Josephson junction \cite{Park2015}.

Instead of coupling Majorana fermions directly to current leads, one can also employ an additional coupling to other bound states formed e.g. in 
quantum dots (QDs) \cite{Liu2011, Leijnse2011, Vernek2014}. The setup proposed in Ref.~\cite{Leijnse2011}  with a Majorana wire tunnel coupled to a QD that is further coupled to a metal lead has very recently also been realized in experiments \cite{Deng1557}. 

Here, we study a setup related to the one used in Ref.~\cite{Deng1557}, but with the important difference that the Majorana wire lies in between
the QD and the current lead with no direct coupling between the QD and the lead, see Fig.~1. We show that this setup exhibits Fano resonance (FR) 
line shapes in conductance (and noise) as a function of the QD level energy $\varepsilon_\text{D}$ with an asymmetry parameter that can change sign as a function
of the bias voltage $eV$ when $eV$ is tuned through the Majorana hybridization energy $2\varepsilon$. We derive analytical formulas for the resonance 
energy $E_R$ and the width $\Gamma_F$ of a corresponding Fano-Beutler formula \cite{Miroshnichenko2010}. Most strikingly, the FR lines are always mirrored 
when the sign of $\varepsilon_\text{D}$ is reversed as long as the QD and the lead are only tunnel-coupled to the nearest MBS. This is a direct sign of the particle-hole symmetry of the zero-energy 
MBS.

However, when the tunnel-coupling to the distant MBS is also finite (see Fig.~3), the symmetry of the conductance (and noise) 
in $\varepsilon_\text{D}$ is destroyed, which is a feature that distinguishes between a MBS and a usual fermionic bound state that can be viewed as composed of two MBS. In the latter case, the 
line shapes of the former mirror-imaged FRs become asymmetric in height, and also can change the sign of the asymmetry parameter. We compare the effective low-energy
model with a numerical treatment of a Kitaev chain and find good agreement at low energies. In addition, we also investigate effects of finite temperatures accessible
in experiments and find that the main features are still well resolved.

We note that Fano resonances in transport have been known for a long time \cite{Miroshnichenko2010} and appear if a resonant transport path interferes with a structureless
transport path. Here, the two transport paths for the injection of a Cooper pair into the grounded superconductor consist of a direct Andreev reflection via the two MBS
and a resonant path where the incoming electron first traverses virtually the quantum wire, visits the QD and then enters the superconducting condensate by reflecting back a hole
to the lead. The interference of the two paths can be tuned via the QD level energy. FRs have been discussed previously in the context of Majorana fermions, but only
in setups where the QD is coupled directly to the lead, either as a function of bias voltage \cite{Dessotti2014,Xia2015,Baranski2017,Xiong2016} and/or flux through a loop containing the MBS \cite{Ueda2014,Jiang2015,Zeng2015} and {\it not} as a function of the QD level energy. Also, the Fano-form of the resonances has been concluded based on a fitting of the conductance
line shapes \cite{Xia2015}, whereas we give analytical expressions for the resonance energy, width and asymmetry of the FR in terms of the model parameters.

The rest of the paper is organized as follows: in Section II, we present the effective low-energy model of the system depicted in Fig.~1(a), present the full counting statistics
results for the cumulant generating function (a derivation is given in Appendix A) and the resulting differential conductance and differential noise. In Section III, we discuss
the regime of appearance and form of the FRs as a function of the QD level energy. In Section IV, we introduce a more general setup including also tunnel-couplings to
the more distant MBS (see Fig.~3(a)). In Section V, we numerically investigate a Kitaev chain tunnel-coupled to a lead on one side and to a QD on the other, including also 
possible non-local tunneling amplitudes from the lead to the Kitaev chain and from the QD to the Kitaev chain. In Section VI we give our conclusion. Appendix A gives a derivation of the
cumulant generating function in terms of Keldysh Green's functions for the setup and the lead. Appendix B discusses the bias voltage dependence of the conductance for different temperatures.

\section{II. Setup and effective model}
\begin{figure}
 \centering
 \subfigure[]{\label{setup1}\def\svgwidth{0.45\textwidth}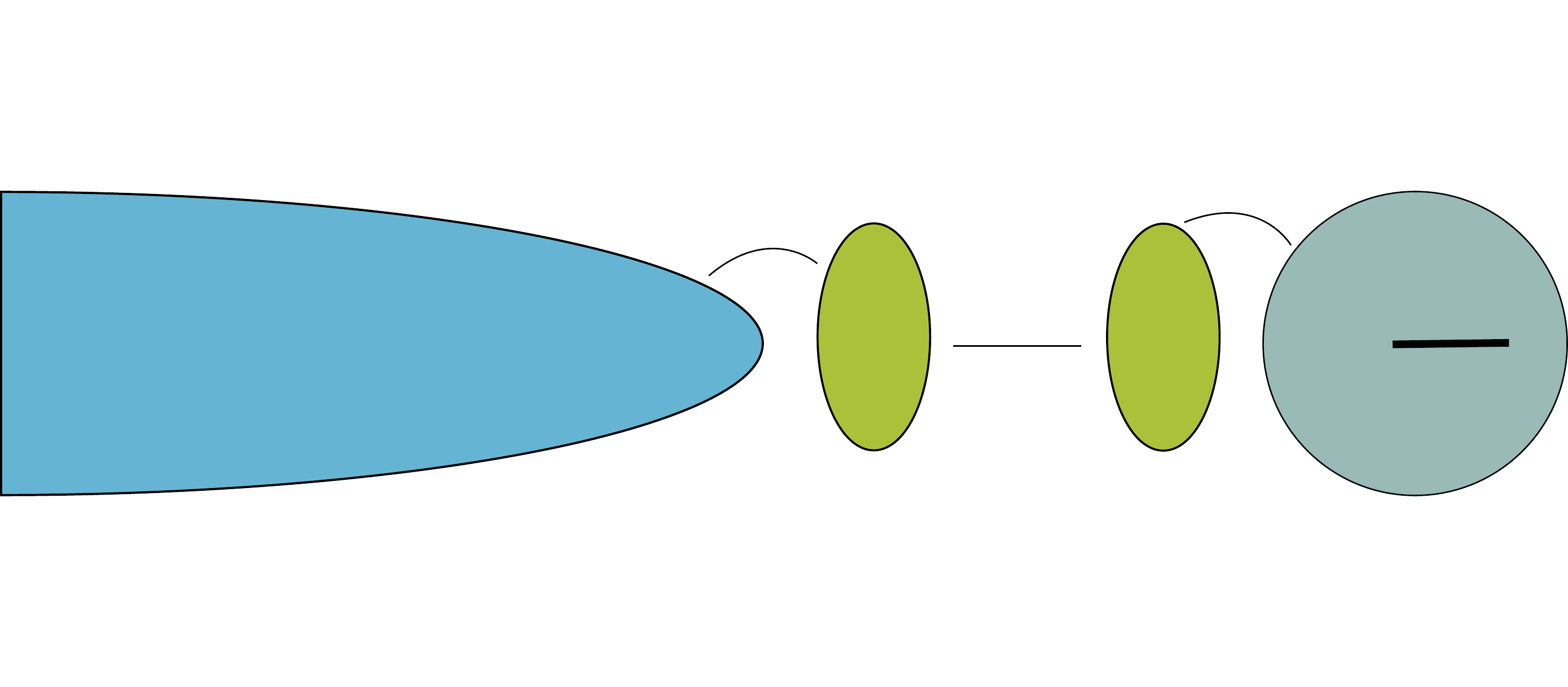}\\
 \subfigure[]{\label{Resandant}
\def\svgwidth{0.46\textwidth}
   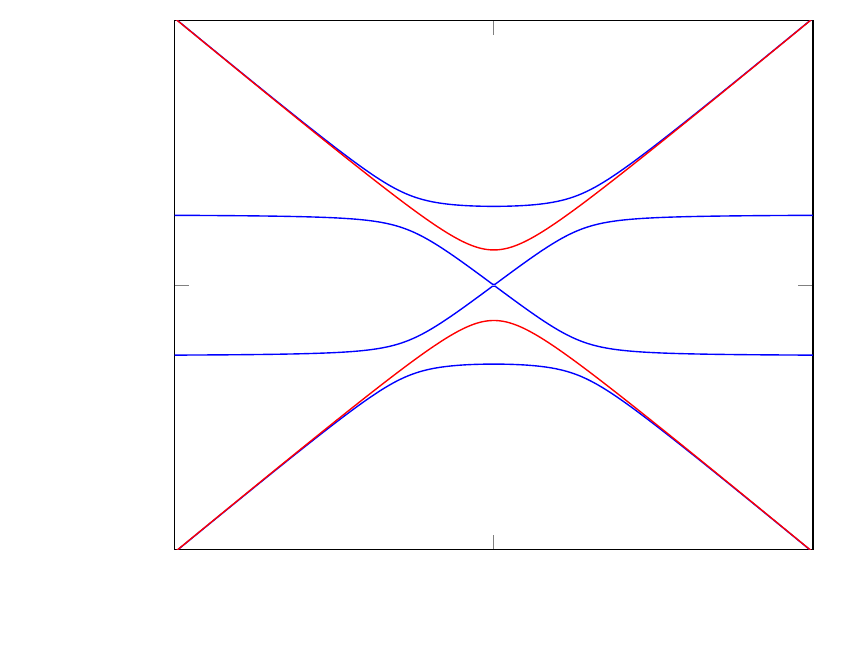
 }
 \caption{(a) Schematic sketch of the considered setup, in which two Majorana bound states $\gamma_1$ and $\gamma_2$ arise in a Kitaev chain. One of them is contacted by a normal conducting lead and the other one
 with a quantum dot )level energy $\varepsilon_{\rm D}$. The two Majorana bound states experience a splitting $\varepsilon$ due to a wave function overlap of the two Majorana fermions. (b) Resonances ($p=1$, blue) and anti resonances ($p=0$, red) of the differential conductance in dependence of the bias voltage $eV$ between lead and the Kitaev chain (grounded), and the
 quantum dot level energy for the parameters $\varepsilon=0.05\Gamma$ and $t_2=0.025\Gamma$. The resonances correspond to the spectrum of the system without the lead. The levels show an avoided crossing if $\varepsilon_{\rm D}\approx \pm 2\varepsilon$}
 \label{Fig1}
\end{figure}

 We consider a setup in which a topologically non-trivial Majorana wire is contacted by a normal conducting lead at one end and  tunnel coupled to a QD at the other end, see Fig.~\ref{Fig1}(a).
 The Majorana edge states of such a wire experience an energy splitting due to their finite wave function overlap $\varepsilon\propto e^{-L/\xi_S}$, where L is the length of the Majorana wire and $\xi_S$ is the 
 superconducting coherence length. We employ a spinless effective description of the MBS and approximate the QD as a single
 fermionic level. The Hamiltonian describing our setup is then given by
\begin{align}
H&=H_L+H_T+H_M+H_{Dot}+H_{TDot}\\
H_M&=i\varepsilon\gamma_1\gamma_2\\
H_{Dot}&=\varepsilon_{\text{D}} d^\dagger d\\
H_{T}&=i\gamma_1\left[t_1\psi^\dagger(0)+t_1^*\psi(0)\right]\\
H_{L}&=-iv_f\int dx \psi^\dagger(x)\partial_x\psi(x)\\
H_{TDot}&=i\gamma_2\left[t_{2}d^\dagger+t_{2}^*d\right],
\end{align}
where $v_f$ is the Fermi velocity in the normal conducting lead, $t_1$ is the tunneling amplitude between the  the Majorana wire and the lead, $t_2$ is the tunneling amplitude
between the Majorana wire and the quantum dot and $\varepsilon_{\text{D}}$ is the energy of the QD level. Here $\gamma_i$ are the Majorana fermion creation operators,
$d$ is the annihilation operator for the quantum dot and $\psi(x)$ is the annihilation operator inside the lead. 
%%%%%%%%%%%%%%%%%%%%%%%%%%%%%%%%%%%%%%%%%%%%%%%%%%%%%%%%%%%%%%%%%%%%%%%%%%%%%%%%%%%%%%%%%%%%%%%%%%%%%%%%%%%%
In order to obtain the transport properties of this setup we use the method of full counting statistics (FCS). The central element of the FCS is the so called cumulant generating function (CGF)~\cite{Levitov2004}, which we calculate
following the Keldysh technique calculations described in Ref.~\cite{Weithofer2014} and summarized in Appendix A. We calculate the CGF in terms of the Keldysh Majorana Green's function
\begin{align}
 \ln\chi(\lambda)&=\frac{\mathcal{T}}{2}\int\frac{d\omega}{2\pi}\ln\big[1+p(\omega)(e^{-2i\lambda}-1)n(\omega)n(-\omega)\notag\\
		 &+p(\omega)(e^{2i\lambda}-1)(n(\omega)-1)(n(-\omega)-1)\big],
		 \label{cgfcalc}
\end{align}
where $n(\omega)=\frac{1}{1+e^{\beta(\omega-eV)}}$ is the Fermi function in the lead with $\beta=1/k_\text{B}T$ the inverse temperature and ${\mathcal{T}}$ is a long measurement time. Andreev reflection (which transports 2 electrons) at energy $\omega$ occurs with probability $p(\omega)$ and corresponds to the only transport channel. The current and the symmetrized zero-frequency noise in the lead can be calculated by taking the first and the second derivative, respectively of the CGF with respect to the counting field $\lambda$.
At zero temperature, the differential conductance and differential noise can be presented analytically as
\begin{align}
 \frac{d I}{dV}&=\frac{d}{d V}\frac{i}{\mathcal{T}}\frac{\partial}{\partial \lambda}\ln\chi\bigg|_{\lambda=0}=\frac{2e^2}{h}p,\\
 \frac{d P}{dV}&=\frac{d}{d V}\frac{-1}{\mathcal{T}}\frac{\partial^2}{\partial \lambda^2}\ln\chi\bigg|_{\lambda=0}=\frac{4e^3}{h}p(1-p),
\end{align}
where
\begin{align}
p=\frac{4\Gamma^2(eV)^2}{\left(\frac{4\varepsilon^2(\varepsilon_{\text{D}}^2-\left(eV\right)^2)}{4|t_2|^2+\varepsilon_{\text{D}}^2-(eV)^2}-(eV)^2\right)^2+4\Gamma^2(eV)^2}\label{cgfmitdot},
\end{align}
with $\Gamma=2\pi\rho_0|t_1|^2$ the tunneling rate between the lead and the nearest MBS ($\gamma_1$) and where $eV$ is the bias voltage between the lead and the grounded topological superconductor hosting the two MBS.
As we are mostly interested in the  dependence of the transport on the QD level energy $\varepsilon_\text{D}$ it is convenient to rewrite
\begin{align}
 p(\varepsilon_\text{D})=\frac{1}{1+\frac{q^2(\varepsilon_\text{D}^2-\varepsilon_{\text{D},\text{max}}^2)^2}{(\varepsilon_\text{D}^2-\varepsilon_{\text{D,0}}^2)^2}}\label{pe},
\end{align}
with
\begin{align}
 \varepsilon_{\text{D},\text{max}}^2&=(eV)^2+\frac{4|t_2|^2\left(eV\right)^2}{(2\varepsilon)^2-(eV)^2},\notag\\
 \varepsilon_{\text{D,0}}^2&=(eV)^2-4|t_2|^2\label{epsilons},\notag\\
 q^2&=\frac{(\varepsilon^2-(eV/2)^2)^2}{\Gamma^2(eV/2)^2}.
\end{align}
Here, $\varepsilon_\text{D}=\varepsilon_{\text{D},\text{max}}$ corresponds to a resonance ($p=1$), at which every incoming electron is Andreev reflected as a hole and $\varepsilon_\text{D}=\varepsilon_{\text{D,0}}$ corresponds to an
anti resonance ($p=0$), where no Andreev reflection is allowed, see Fig.~\ref{Fig1}(b). The parameter $q^2$ characterizes the Andreev reflection probability without the coupling to the QD. The special point with
$\varepsilon_\text{D}=eV=0$ is discussed in Appendix~\ref{secZBP}B.
\section{III. Fano Resonances}
Fano resonances are characterized by a typical asymmetric line shape~\cite{Fano1935} and appear when a continuous path interferes with a discrete path. 
In the considered setup, the two paths correspond to a direct Andreev reflection without including the QD, and a path where the incoming electron first traverses the quantum wire, visits the QD and returns to the quantum wire where 
the electron forms a Cooper pair by emission of a hole into the lead. The second process depends on the QD level energy and corresponds to the discrete path. The interference of both processes leads to a resonance as a function of $\varepsilon_{\rm D}$.

We intend to describe this resonance in the Andreev reflection probability with the Fano-Beutler formula~\cite{Miroshnichenko2010}
\begin{equation}
 p_{\rm FB}(\varepsilon_\text{D})=\frac{1}{1+q^2}\frac{((\varepsilon_{\text{D}}-E_R)/(2\Gamma_F)+q)^2}{1+((\varepsilon_{\text{D}}-E_R)/(2\Gamma_F))^2}\label{Fanoformula},
\end{equation}
where $\Gamma_F$ is the width of the FR and $E_R$ is the resonance energy. The parameter q describes the asymmetry of the resonance and its sign determines whether the destructive 
or constructive interference can be found at smaller
energies. It is important to note that this formula only describes one resonance, while we always find two resonances in the considered setup as we will discuss now.

According to Eq.~(\ref{Fanoformula}), the resonance $(p_{\rm FB}(\varepsilon_\text{D})=1)$ and the anti-resonance ($p_{\rm FB}(\varepsilon_\text{D})=0$) of a FR as a function of $\varepsilon_{\text{D}}$ apear at
\begin{align}
 \varepsilon_{\rm D}=\varepsilon_{\text{D},\text{max}}&\equiv2\Gamma_F/q+E_R,
 \label{Fanoparameter1}
 \end{align}
 and at
 \begin{align}
 \varepsilon_{\rm D}=\varepsilon_{\text{D},0}&\equiv-2q\Gamma_F+E_R\label{Fanoparameter2},
\end{align}
respectively.

Inserting these into Eq.~(\ref{pe}) and considering a small width $|\Gamma_F/E_R|\ll 1$ we can simplify the Andreev reflection probability to
\begin{align}
p(\varepsilon_\text{D})\approx\left(\frac{1}{1+q^2}\right)\ &\left(\frac{(\frac{\varepsilon_\text{D}-E_R}{2\Gamma_F}+q)^2}{1+(\frac{\varepsilon_\text{D}-E_R}{2\Gamma_F})^2}\right)\notag\\ 
\times&\left(\frac{(\frac{\varepsilon_\text{D}+E_R}{2\Gamma_F}-q)^2}{1+(\frac{\varepsilon_\text{D}+E_R}{2\Gamma_F})^2}\right).
\end{align}
This describes a product of two Fano-Beutler formulas. These two resulting FRs have opposite asymmetry parameters $q$ and opposite resonance energies $E_R$.  We further approximate the Andreev reflection probability in the limit of large positive $E_R/\Gamma_F$ compared to $q$ with 
\begin{equation}
p(\varepsilon_\text{D})\approx\frac{1}{1+q^2}\frac{(\frac{|\varepsilon_\text{D}|-E_R}{2\Gamma_F}+q)^2}{1+(\frac{|\varepsilon_\text{D}|-E_R}{2\Gamma_F})^2}.      
\end{equation}
This equation describes two FRs that are mirrored at $\varepsilon_\text{D}=0$ and which have opposite signs of $q$ and of $E_R$.
Note that the asymmetry parameter $q=\pm((eV/2)^2-\varepsilon^2)/(\Gamma(eV/2))$ changes sign at both FRs when the bias voltage is tuned through the MBS hybridization energy $2\varepsilon$ (cf. Fig.~\ref{Resandant}). Here the $+$ corresponds
to the FR at positive $\varepsilon_\text{D}$, while $-$ refers to the FR at negative dot level energy.
\begin{figure}[t]
 \centering
 \subfigure[]{\label{resandtii}

\includegraphics[width=0.46\textwidth]{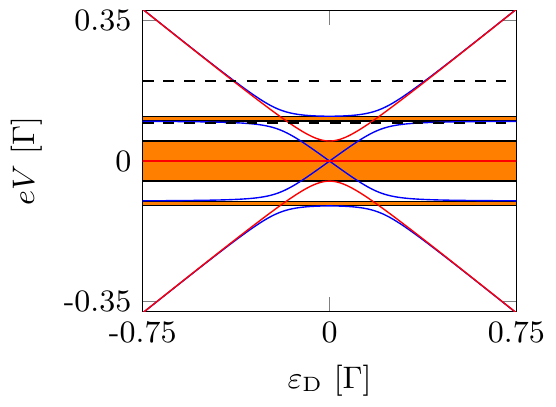}
 }
 \subfigure[]{\label{fanores}

\includegraphics{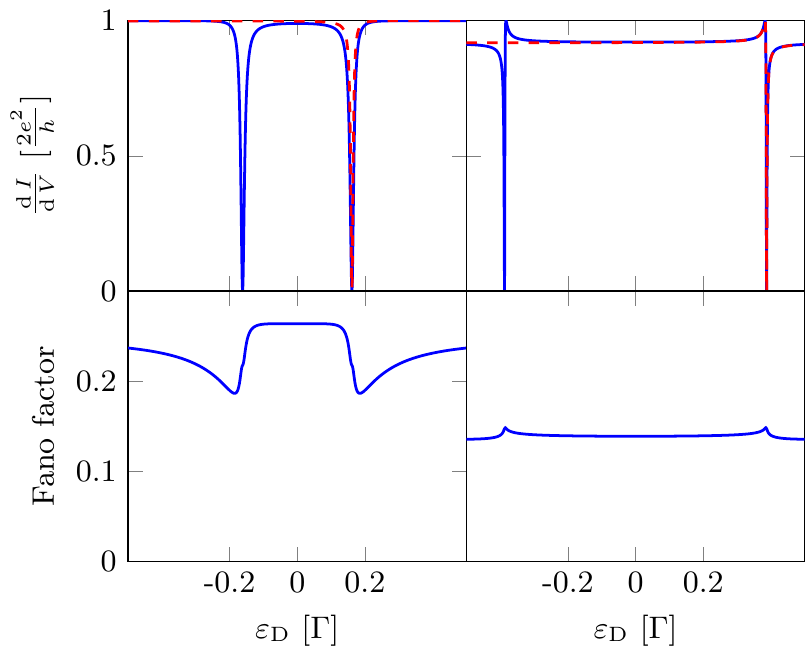}
}
 \caption{(a) Resonances and anti resonances in the setup with $\varepsilon=0.05\Gamma$ and $t_2=0.025\Gamma$. The dashed lines correspond to the bias voltages chosen for the plots in (b) ($eV=0.095\Gamma$ for the left plots, $eV=0.2\Gamma$ for the right plots). Orange background color marks the parameter
 space in which no Fano resonances arise. (b) Differential conductance (top) or Fano factor (bottom) vs. dot level for different bias voltages between lead and topological superconductor. The blue line is the exact calculated differential conductance (Fano factor), whereas the red dashed line
 is the approximation using Eq.(\ref{Fanoformula}), where $E_R$ and $\Gamma_F$ are given by Eq.~(\ref{ER}) and Eq.~(\ref{GF}), respectively.}
\end{figure}
%%%%%%%%%%%%%%%%%%%%%%%%%%%%%%%%%%%%%%%%%%%%%%%%%%%%%%%%%%%%%%%%%%%%%%%%%%%%%%%%%%%%%%%%%%%%%%%%
In order to represent the parameters of the Fano-Beutler formula with the parameters of the microscopic model we insert $\varepsilon_{\text{D},\text{max}}$ and $\varepsilon_{\text{D},0}$ from Eq.~(\ref{Fanoparameter1}) and Eq.~(\ref{Fanoparameter2}) into Eqs.~(\ref{epsilons}) and focus on the FR at positive dot level energy

\begin{align}
 E_R&=2\frac{ \left(\frac{eV}{2}\right)^2 \Gamma ^2 \sqrt{\left(\frac{eV}{2}\right)^2-t_2^2}}{ \left(\frac{eV}{2}\right)^2 \Gamma^2+\left(\varepsilon^2-\left(\frac{eV}{2}\right)^2\right)^2}\label{ER},\\
 &+2\frac{ (\left(\frac{eV}{2}\right)^2-\varepsilon^2)^{3/2} \sqrt{ \left(\frac{eV}{2}\right)^2\left(\left(\frac{eV}{2}\right)^2-\varepsilon^2-|t_2|^2\right)}}{ \left(\frac{eV}{2}\right)^2 \Gamma^2+\left(\varepsilon^2-\left(\frac{eV}{2}\right)^2\right)^2}\notag\\
 \Gamma_F&=\frac{\left(\left(\frac{eV}{2}\right)^2-\varepsilon^2\right) \sqrt{\left(\frac{eV}{2}\right)^2-t_2^2}}{ \left(\frac{eV}{2}\right) \Gamma+\left(\varepsilon^2-\left(\frac{eV}{2}\right)^2\right)^2/ \left(\left(\frac{eV}{2}\right) \Gamma\right)}\label{GF}\\
 &-\frac{\sqrt{\left(\frac{eV}{2}\right)^2-\varepsilon^2} \sqrt{\left(\frac{eV}{2}\right)^2 \left(\left(\frac{eV}{2}\right)^2-\varepsilon^2-t_2^2\right)}}{ \left(\frac{eV}{2}\right) \Gamma+\left(\varepsilon^2-\left(\frac{eV}{2}\right)^2\right)^2/ \left(\left(\frac{eV}{2}\right) \Gamma\right)}\notag.
\end{align}

FRs can not be found in a regime where $(eV/2)^2<|t_2|^2$. In this regime $E_R$ and $\Gamma_F$ would become complex numbers whereas the differential conductance has to be real valued. Therefore, the description with the Fano formula fails in that regime.
Also, in the regime where $\varepsilon<|eV|/2<\sqrt{|t_2|^2+\varepsilon^2}$ a description with the Fano-Beutler formula is not possible (cf. Fig.~\ref{resandtii}).\\
The assumption of a large resonance energy $E_R$ and small width $\Gamma_F$ can now be represented with the model parameters and it can be seen that the description with the Fano formula holds for 
$|t_2|\ll |eV|$. This corresponds to a weakly coupled QD. This prediction can also be seen in Fig.~\ref{fanores}, because for a higher bias voltage between lead and topological superconductor the approximation with the
FR line shape becomes better. 

Fig.~\ref{resandtii} highlights the regimes where no FRs appear (orange boxes). In these regimes exclusively either a resonance or anti resonance is crossed when changing $\varepsilon_\text{D}$ at fixed bias voltage $eV$. The dashed horizontal lines denote different bias voltages used for the plots in Fig.~\ref{fanores}. The differential conductance traces are in very good agreement with the deduced Fano form. The dashed lines result from the Fano-Beutler formula Eq.~(\ref{Fanoformula}) with the parameters $E_R$ and $\Gamma_F$ expressed with the microscopic parameters of our model via Eq.~(\ref{ER}) and Eq.~(\ref{GF}). Note that only one resonance is described by the Fano-Beutler formula, the other one is the mirror image with respect to the line $\varepsilon_{\rm D}=0$. The change of the sign of q as a function of the bias voltage gives a direct measure for the overlap energy $\varepsilon$. When $eV=2\varepsilon$ ($q=0$), the FR line shapes become symmetric. This results in an experimental signature to obtain the overlap energy $\varepsilon$. Note that the conductance is always quantized at resonances to $2e^2/h$ and is {\it symmetric} in $\varepsilon_D$. These features are clear signatures of tunneling into {\it single} MBS, which are zero energy- and particle-hole symmetric-states. The symmetry in the bias voltage (not shown) is more fundamental as it is true for all superconductors. In Fig.~\ref{fanores} we in addition show the Fano factor $F=P(V)/eI(V)$ which also exhibits features at the positions of the FRs in conductance. For the chosen parameters in the plots, the noise is always sub-poissonian $(F<1)$ but does not vanish at the conductance resonances which is a sign of a strong energy dependence of the Andreev reflection. In the next section, we will demonstrate that the symmetry in $\varepsilon_\text{D}$ is waived when both MBS couple directly to the QD. 

\section{IV. Non-Majorana Fermionic Coupling}
\begin{figure}
\centering
\subfigure[]{\label{generali}\def\svgwidth{0.45\textwidth}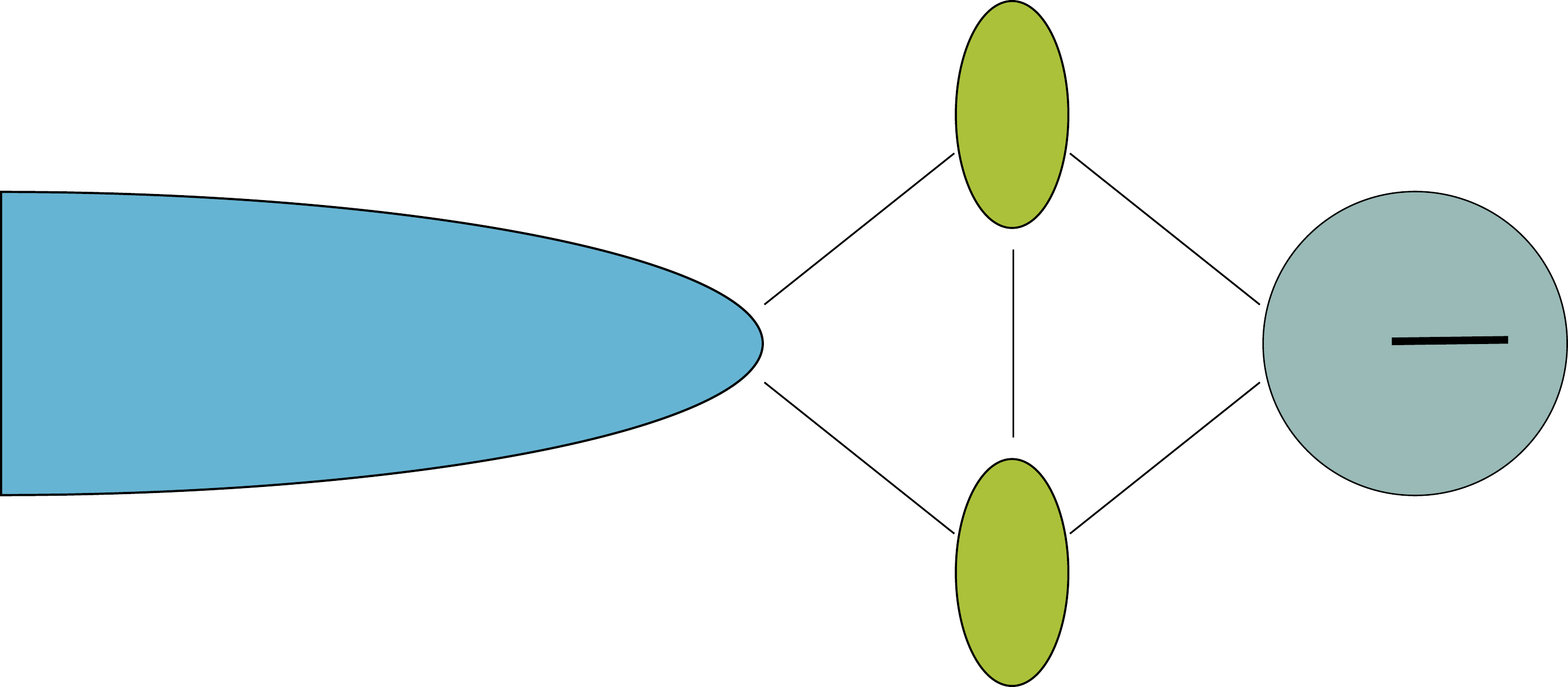}\quad
\subfigure[]{\label{fermi}

\includegraphics{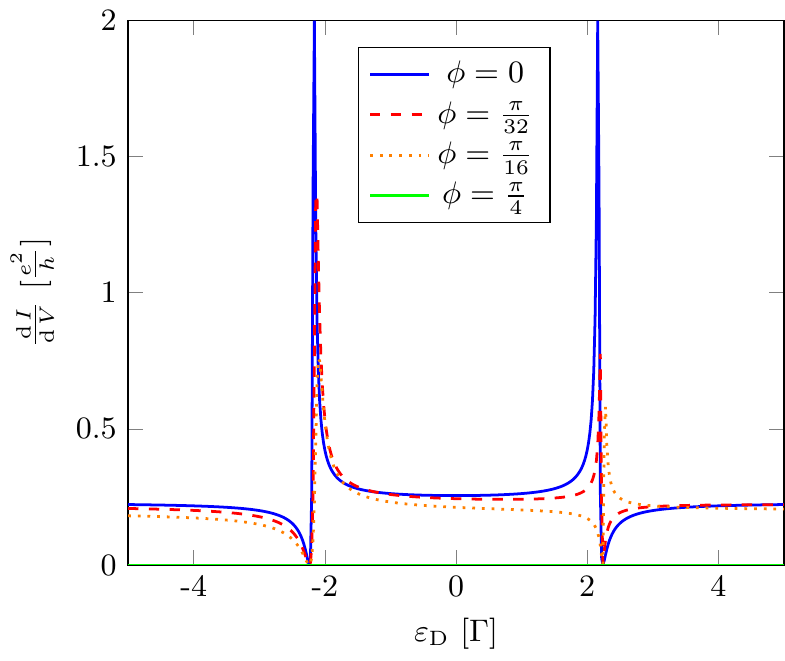}}
\caption{(a) Sketch of the extended setup. By varying the parameter $\phi$ we can tune the system from pure Majorana coupling ($\phi=0$) to pure Dirac fermionic coupling ($\phi=\frac{\pi}{4}$),
in order to find unique signatures of pure Majorana fermionic coupling. (b) Differential conductance in the extended setup for different $\phi$. In the Majorana coupling case $\phi=0$ the Fano resonances
are quantized and are mirrored at $\varepsilon_\text{D}=0$ due to the real valued properties of the Majorana fermion. For $0<\phi<\frac{\pi}{4}$ the resonances are no longer quantized and the peaks are no
longer symmetric. Parameters are $eV=3\Gamma$, $|t_2|=\Gamma$, $\varepsilon=0.4\Gamma$.
}
\end{figure}

In this section we compare
the FRs which arise from an effective model with non-Majorana fermionic couplings, see Fig.~\ref{generali}.\\
For this, we consider a more general setup in which the normal conducting lead does not only contact one Majorana bound state, but also the second Majorana fermion. The same is true for the connection to the QD.
A schematic sketch of this setup can be seen in Figure~\ref{generali}.
A physical realization of this could be a very short chain, so that the lead not only contacts the first, but also (however weaker) the last site.
The Hamiltonian of this effective setup is
\begin{align}
 &H=H_L+i\varepsilon\gamma_1\gamma_2+\varepsilon_\text{D}d^\dagger d\\
 &-i\gamma_1\big[t_1\cos(\phi)\psi+t_1\cos(\phi)\psi^\dagger\notag\\
 &+i t_2\sin(\phi)d-it_2\sin(\phi)d^\dagger\big]\notag\\
&-i\gamma_2\big[it_1\sin(\phi)\psi-it_1\sin(\phi)\psi^\dagger\notag\\
&+ t_2\cos(\phi)d+t_2\cos(\phi)d^\dagger\big]\notag,
 \end{align}
where $H_L$, $\gamma_i$ and $\psi$ are the same operators as before and $t_1$ and $t_2$ are real valued. In this setup, we can tune between a Majorana-like coupling or a Dirac fermionic coupling by tuning the parameter $\phi$, where
$\phi=0$ is the Majorana case as discussed above and  $\phi=\frac{\pi}{4}$ corresponds to the case of a single Dirac fermionic site. In between, we have a mixture of Majorana fermionic and Dirac fermionic coupling.
The CGF can now be calculated using the same formalism as before and is formally the same as in Eq.(\ref{cgfcalc}), however the Andreev reflection probability $p(\omega)$ is different.

\begin{figure}
\centering

\includegraphics[width=0.5\textwidth]{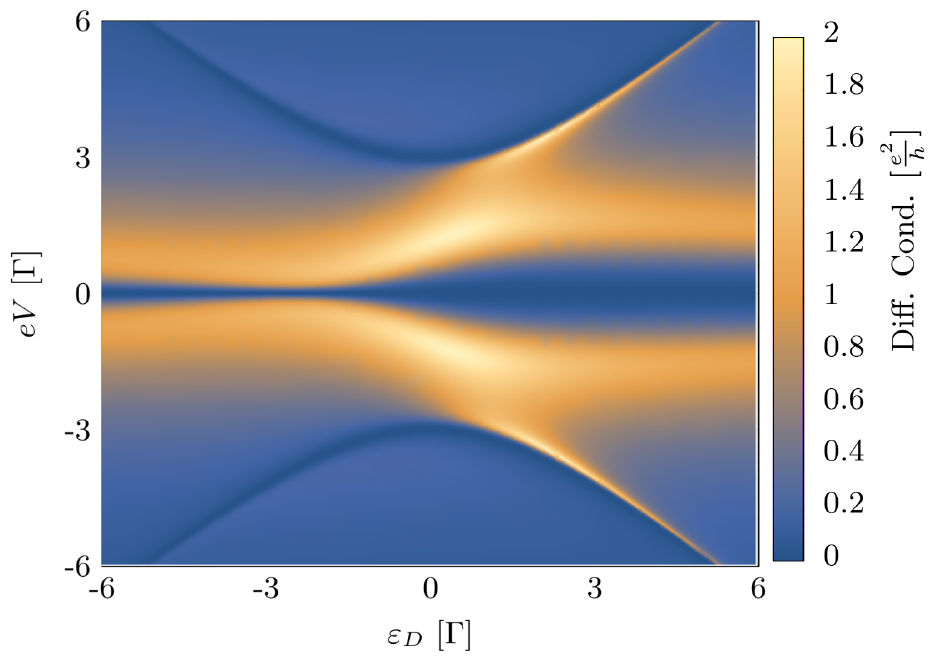}
 \caption{(a)Differential conductance as function of dot level energy and bias voltage for non-Majorana fermionic couplings with $\phi=0.4$, $\varepsilon=0.5\Gamma$, $|t_2|=1.5\Gamma$. The maxima of the differential conductance correspond to the eigenenergies of the system without the lead. }
 \label{spektra}
\end{figure}

By taking the derivative of the CGF with respect to $\lambda$ we can calculate the differential conductance.
It is shown in Figure~\ref{fermi} as function of the QD level energy at fixed bias voltage for different $\phi$. For $\phi=0$ we reproduce the results which we have discussed before, while in
the $\phi=\frac{\pi}{4}$ case the transport into the superconductor is completely blocked. The case of $0<\phi<\frac{\pi}{4}$ is more interesting for us. We can clearly see that the resonance peaks of the 
two arising FRs are no longer quantized and even have different heights for $\text{sign}(\varepsilon_\text{D})=\pm1$. We can further tune the system into a regime where the signs of the asymmetry 
parameters of the two FRs are no longer opposite as is always the case in the Majorana fermionic coupling limit ($\phi=0$). This is due to the fact that a non-Majorana fermionic level no longer couples to electrons
and holes in the same way, which results in a different behaviour for positive and negative dot level energies. In Fig.~\ref{spektra} we present the differential conductance as a function of $\varepsilon_{\rm D}$ and $eV$ and for $\phi=0.4$. We clearly see that the conductance traces become asymmetric in the QD level energy and that the maxima are not generally quantized to $2e^2/h$ anymore.

Therefore, a unique signature of MBS is that the two FRs are always mirrored as a function of $\varepsilon_\text{D}$ at the ordinate, when the coupling to the leads is "local" like in the setup of Fig.~\ref{Fig1}. This can be explained with the real-valued nature of the Majorana bound states. Because the MBS are described by real-valued spinors, they couple to electron- and hole-like excitations in the same way and $\frac{dI}{dV}(\varepsilon_\text{D})=\frac{dI}{dV}(-\varepsilon_\text{D})$. This result is even valid for finite temperatures and in more realistic setups as we will show explicitly in the next section.
%%%%%%%%%%%%%%%%%%%%%%%%%%%%%%%%%%%%%%%%%%%%%%%%%%%%%%%%%%%%%%%%%%%%%%
\section{V. Kitaev-Chain Model}
The next step towards an experiment is to consider
a one dimensional $p$-wave superconductor and to discretize its Hamiltonian on a one-dimensional lattice
\begin{align}
 &H_{p\text{-wave}}=\frac{1}{2}\int dx\left(\begin{smallmatrix}\Psi^{\dagger}(x),&\Psi(x)\end{smallmatrix}\right)\left(\begin{smallmatrix}
                                                                                  -\tilde t\partial_x^2+U&-i2\Delta \partial_x\\
                                                                                  -i2\Delta^* \partial_x&\tilde t\partial_x^2-U
                                                                                 \end{smallmatrix}
\right)\left(\begin{smallmatrix}
              \Psi(x)\\
              \Psi^\dagger(x)
             \end{smallmatrix}
\right)\notag\\
&=\sum_{j=1}^N \left(U-\frac{2{\tilde t}}{a^2}\right)
                  \left(c_j^\dagger c_j-\frac{1}{2}\right)\notag\\
                  &-\sum_{j=1}^{N-1}\left(\frac{\tilde t}{a^2}(c_j^\dagger c_{j+1}+c_{j+1}^\dagger c_j)+\frac{i{\Delta}}{a}c_j^\dagger c_{j+1}^\dagger+\frac{i{\Delta}^*}{a}c_jc_{j+1}\right)\notag\\
&=\frac{i}{2}\sum_{j=1}^{N}\left(U-\frac{2\tilde t}{a^2}\right)\gamma_{2j-1}\gamma_{2j}+\frac{i}{2}\sum_{j=1}^{N-1}\left[\left(\frac{\tilde t}{a^2}+\frac{|\Delta|}{a}\right)\gamma_{2j}\gamma_{2j+1}\right.\notag\\
&\left.+\left(-\frac{{\tilde t}}{a^2}+\frac{|\Delta|}{a}\right)\gamma_{2j-1}\gamma_{2j+2}\right],
\end{align}
where $\tilde t=\hbar^2/(2m^*)$ and in last step we followed a transformation to Majorana fermions similar as described in Ref.~\cite{Kitaev2001}. $U$ plays the role of a chemical potential and can be used to tune the topological properties of the chain \cite{Kitaev2001}. With this Kitaev chain Hamiltonian we can now increase the number of sites while keeping the length $L=Na$ of the wire fixed.\\
We now couple, with tunneling-amplitude $t_2$, a single Dirac-fermionic site (the QD) with energy $\varepsilon_{\rm D}$ and annihilation operator $d$ to the last Dirac-fermionic site ($c^{\dagger}_{N}$) of the chain, and also tunnel-couple the first Dirac-fermionic site of the Kitaev chain ($c_1$) to the Fermi lead ($\psi^{\dagger}(0)$) with amplitude $t_1$. 
The total Hamiltonian written in terms of Majorana fermions then reads
\begin{align}
 &H=H_{p\text{-wave}}+H_{\text{L}}\\
 &+\frac{i}{2}\left[-t_2\gamma_{2N}\gamma_{2N+1}+t_2\gamma_{2N-1}\gamma_{2N+2}+\varepsilon_{\text{D}}\gamma_{2N+1}\gamma_{2N+2}\right]\notag\\
 &+\frac{i}{2}\gamma_1\left[(it_1)^{*}\psi(0)+(it_1)\psi^{\dagger}(0)\right]-\frac{i}{2}\gamma_2\left[t_1\psi^{\dagger}(0)+t_1^*\psi(0)\right]\notag,
 \end{align}
 where we included the QD with the Hamiltonian
 \begin{align}
  H_\text{Dot}=\varepsilon_\text{D}\left(d^\dagger d-\frac{1}{2}\right)=\frac{i}{2}\varepsilon_{\text{D}}\gamma_{2N+1}\gamma_{2N+2}.
 \end{align}
Without loss of generality, we have chosen $t_2$ to be real. 
We implement this Hamiltonian in our FCS formalism and solve the formula for the CGF numerically using Eq.~(\ref{levitov}).
\begin{figure}[h!]
\centering

  \subfigure[]{\label{Num}\def\svgwidth{0.45\textwidth}
  \input{Numericplot.pdf_tex}}\\
  \subfigure[]{\label{Temp}
 \includegraphics[]{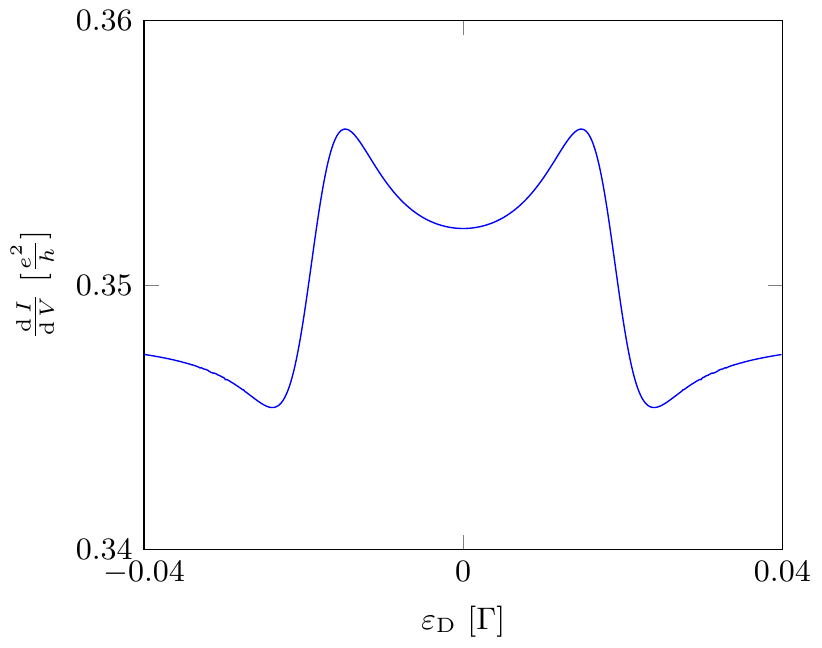}}
  \caption{
   (a) Differential conductance as function of the dot level energy and the bias voltage between lead and the Kitaev chain with $N=100$ sites. The parameters are $a=6\cdot10^{-4}$, 
   $\tilde t=1/4500$, $|\Delta|=1/700$, $\Gamma=1$, $U=0.2$ and $t_2=0.09$, which corresponds to the topologically non-trivial phase. The $p$-wave gap is around $\Delta_p=0.1\Gamma$ and above this
   gap the differential conductance is on the order of $10^{-3}\frac{e^2}{h}$. At lower energies resonances can be seen which correspond to in gap states. The strongest resonance is the Majorana
   induced resonance with a quantized differential conductance of $\frac{2e^2}{h}$. The influence of the dot can be seen best at the Majorana resonance and results in destructive interference.
   The inset shows the low energy section. The quantized resonances corresponding to the dot level can not be clearly seen because of resolution problems. (b) Fano resonance at finite temperature $k_{\rm B}T=0.01\Delta_p$. This temperature is in agreement with comparable experiments~\cite{Deng1557}. The symmetry with respect to the dot level energy is conserved also at finite temperatures.
  }
\end{figure}
The resulting differential conductance is shown in Fig.~\ref{Num}. In the low-energy limit (inset of Fig.~\ref{Num}) we see a good agreement with the effective model (Fig.~\ref{Resandant}). In the numerical data we see not only the
differential conductance peaks which result from Majorana bound states, but also more states inside the spectral p-wave gap ($\Delta_p=0.1\Gamma$) of the Kitaev chain.\\
Even calculations at finite temperatures show that the symmetry property of the two Fano resonances in the MBS-QD system are conserved as can be seen in Fig.~\ref{Temp}. The results from the effective model calculations can be reproduced perfectly.

\begin{figure}[h!]
\includegraphics{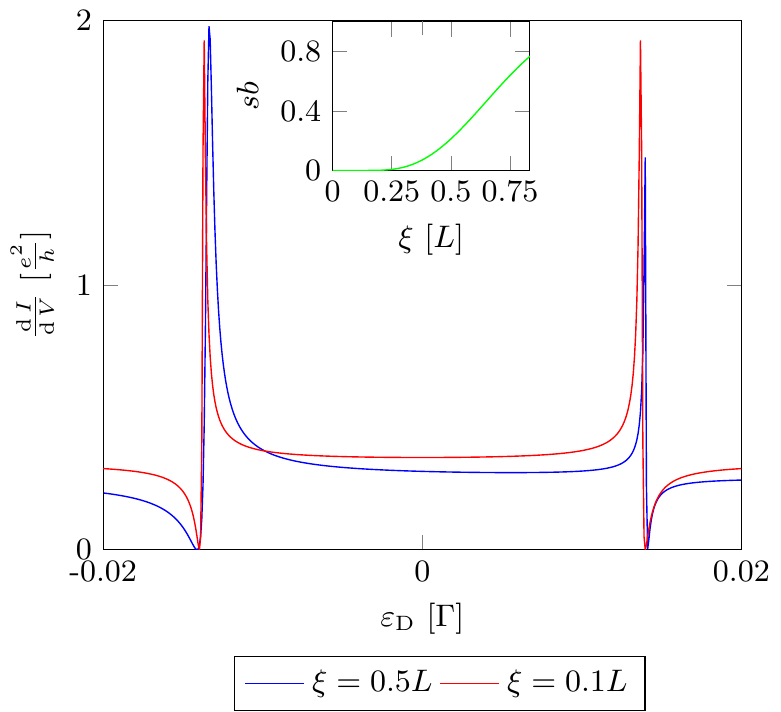}
\caption{Fano resonances in the Kitaev chain for non-trivial couplings to normal conducting lead and QD using $t_2=0.15\Gamma$. The other parameters are the same as in Fig.~\ref{Num}. In the case of long range couplings ($\xi=0.5L$) the two Fano resonances are no longer symmetric. For a shorter decay
length ($\xi=0.1L$) the symmetry of the two FRs is restored. The inlay shows the symmetry breaking parameter $sb=(p(\varepsilon_\text{D}=eV)-p(\varepsilon_\text{D}=-eV))/p(\varepsilon_\text{D}=eV)$ vs. the tunneling decay length.}
\label{nontrivKitaev}
\end{figure}

As a next step we consider again the Kitaev chain and introduce a coupling between the normal conducting Fermi lead and each site and between the QD and each site of the Kitaev chain.
We assume that this tunneling is exponentially suppressed with the length of the tunneling distance. With this modification, the tunneling Hamiltonian becomes 
\begin{align}
 H_\text{T}=\sum_{n=1}^{N}\left( t_1\,e^{-\frac{(n-1)a}{\xi}} \psi^{\dagger}(0) c_n+t_2\,e^{-\frac{(N-n)a}{\xi}}c_{n}^\dagger d +{\rm h.c.}\right),
\end{align}
where $\xi$ is the length scale on which the tunneling amplitude decays. In Fig.~\ref{nontrivKitaev} we present the resulting differential conductance using this tunneling Hamiltonian. When $\xi$ is on the
order of the length of the wire the symmetry of the two FRs is broken. However, for $\xi=0.1L$ the symmetry is restored.
We can explain this behaviour by reconsidering the effective model. Here, the symmetry is broken if the dot couples to both MBS. We consider the coherence length of
the bound states $\xi_S$ which is given in Ref.~\cite{Kitaev2001}. In order to couple to both MBS the tunneling decay length needs to be bigger than the difference between the 
length $L$ of the wire and $\xi_S$
\begin{align}
 L-\xi_S\ll\xi.
\end{align}
The inlay in Fig.~\ref{nontrivKitaev} supports our statement. The symmetry breaking is not visible for all decay length's but starts at a finite value. For the parameters we used in our numerical calculations we find that $\xi_S\approx0.6L$,
which corresponds to the increase in the symmetry breaking parameter
\begin{align}
 sb=\frac{p(\varepsilon_\text{D}=eV)-p(\varepsilon_\text{D}=-eV)}{p(\varepsilon_\text{D}=eV)},
\end{align}
for $\xi=0.5L$ relative to $\xi=0.1L$.
Therefore, the symmetry in $\varepsilon_\text{D}$ can be seen as a signature of the coupling to a single MBS. Because, even if there is a finite overlap energy $\varepsilon$ between the two MBS, and so the ground state is a Dirac-fermionic state, the symmetry in $\varepsilon_D$  vanishes
if and only if both Majorana fermions are contacted.

\section{VII. Conclusion}
In summary, we studied the resonances in differential conductance and shot noise for the normal conducting lead - Majorana wire - QD setup. We showed that these resonances, as a function of the QD level energy $\varepsilon_\text{D}$, come in pairs and
can be described with the Fano-Beutler formula which proofs that the observed line-shapes are Fano resonances.

We investigated two models: First, we used an effective low-energy model for the Majorana wire restricting ourselves to hybridized Majorana end states with overlap energy $\varepsilon$. 
In the case where the lead and the QD are coupled to only one but opposite MBS each, these two Fano resonances are symmetric under $\varepsilon_\text{D}\rightarrow-\varepsilon_\text{D}$ for all ranges of parameters
including finite temperature. At zero temperature, the conductance resonances are always quantized to $2e^2/h$. We also observe a rather striking characteristic that the asymmetry parameter of the Fano resonances changes sign when the bias voltage is tuned through the Majorana fermion overlap energy, a result that is independent on the parameter of the QD giving independent access on the Majorana splitting. In an extended setup where {\it both} Majorana end states are coupled to the lead and QD, the symmetry of the conductance traces as a function of $\varepsilon_\text{D}$ is waived and the resonances at zero temperature are not generally quantized anymore.

We support these findings from the effective model calculation by considering a Kitaev chain with a finite length giving us control over the Majorana overlap numerically and find good agreement with the analytical model in the low-energy sector. In particular, we also find asymmetric Fano Resonances when the tunneling between the chain and the lead and between the chain and the QD becomes long ranged. In this context long means that the tunneling amplitude from the dot and from the lead to the chain needs to reach the MBS at the other side of the chain.

We consider this symmetry in $\varepsilon_\text{D}$ to be a unique signature of the MBS as it is only lifted if we couple the dot/lead to both MBS and therefore to a non-Majorana fermionic bound state. 

We note that the conductance maxima reflecting the spectrum of the MBS-quantum dot system have similarities to a recent experiment in a slightly different setup (not showing Fano resonances), where a quantum dot is coupled to a Fermi lead on one side and a quantum wire on the other \cite{Deng1557}. It would be interesting to investigate the role of possible non-local couplings in this setup which should lead to our predicted asymmetries under sign-inversion of the QD level energy ($\varepsilon_{\rm D}\rightarrow  -\varepsilon_{\rm D}$).

While finishing this manuscript we became aware of two recent related preprints \cite{Clarke2017, Prada2017} that discuss the energy spectrum of a setup similar to the one in Section IV (without the lead) and note an asymmetry of the spectrum as a function of the QD level energy similar to the maxima in our Fig.~4. However, both preprints do not calculate transport properties.

\section{Acknowledgments}
PR would like to thank R\'{a}mon Aguado, Alfredo Levy-Yeyati, Sunghun Park, Elsa Prada, and Pablo San-Jos\'{e} for very helpful discussions on this project close to its finalization. We also thank for financial support from the DFG Grant No.
RE 2978/1-1, from the Lower Saxony PhD-programme "Contacts in Nanosystems", the Research Training
Group GrK1952/1 "Metrology for Complex Nanosystems",
and the Braunschweig International Graduate
School of Metrology B-IGSM.

\appendix
\section{A. Calculation of the CGF}
\label{cmbs}
In order to use the following formalism to obtain the CGF we transform the fermionic creation and annihilation operators in terms of Majorana operators.
So we rewrite
\begin{align}
H_M´&=H_M+H_{Dot}+H_{TDot}\\
&=\frac{i}{2}\sum_{k,l}A_{kl}\gamma_k\gamma_l.
\end{align}
Majorana fermion networks can in general be described by the Hamiltonians
\begin{align}
H&=H_L+H_M+H_T\\
H_M´&=\frac{i}{2}\sum_{k,l}A_{kl}\gamma_k\gamma_l\\
H_L&=-i v_f\sum_\alpha\int dx\psi_\alpha^\dagger(x)\partial_x\psi_\alpha(x)\\
H_T&= -i  \sum_{\alpha\beta} \gamma_\beta \left[ t_{\alpha\beta}\psi_\alpha^\dag(0) + t_{\alpha\beta}^*\psi_\alpha(0) \right],
\end{align}
where $\gamma_k$ is the real valued creation operator for Majorana bound state $k$ and $A_{kl}=A^*_{kl}=-A_{lk}$, which has to be specified for each setup, $H_L$ describing the leads, where $\psi_\alpha$ and $\psi_\alpha^\dagger$ are the fermionic annihilation and creation operators in lead $\alpha$ and
$H_T$ describing the tunnel coupling between the leads and the Majorana fermions. Here $v_f$ is the Fermi velocity and $t_{\alpha\beta}$ is the tunneling amplitude between lead $\alpha$ and Majorana bound state $\beta$.\\
We closely follow the derivation of the cumulant generating function using functional integrals provided in~\cite{Weithofer2014}, which we generalize to arbitrary couplings between normal leads and Majorana fermions and find for the CGF
\begin{equation}
 \ln\chi(\vec\lambda)=\frac{\mathcal{T}}{2}\int\frac{d\omega}{2\pi}\ln\frac{\det[D^\lambda]^{-1}}{\det[D^{\lambda=0}]^{-1}},
 \label{levitov}
\end{equation}
where $D^\lambda(\omega)$ is the Green's function for the Majorana fermions including the influence of the leads and the counting fields. Its inverse is
\begin{align}
 \left[D^\lambda\right]^{-1}(\omega)=\left[D^{(0)}\right]^{-1}(\omega)-\Sigma^\lambda(\omega)
 \label{Dyson}
\end{align}
Here $D^{(0)}$ is the unperturbed Majorana Green's function and $\Sigma^\lambda$ is the self energy containing the counting fields.\\
The unperturbed Majorana Green's function is defined as $[D^{0}]_{\alpha\beta}(t,t')=-i\braket{T_{\mathcal{C}}\gamma_\alpha(t)\gamma_\beta(t')}$ with $T_{\mathcal{C}}$ the time-ordering operator on the Keldysh contour, and can be calculated using the Heisenberg equation of motion for the free Majorana fermion operators
\begin{equation}
 \frac{d\gamma_\alpha}{dt}=i[H_{M},\gamma_\alpha]=\sum_{\beta}2 A_{\alpha\beta}\gamma_\beta,
 \label{HOM}
\end{equation}
where we used that $A$ is a real skew symmetric matrix. As the unique solutions of Eq.~(\ref{HOM}) are given by
\begin{equation}
 \gamma_\alpha(t)=\sum_\beta \mathcal{B}_{\alpha\beta}(t)\gamma_\beta,
\end{equation}
%%%%%%%%%%%%%%%%%%%%%%%%%%%%%%%%%%%%%%%%%%%%%%%%%%%%%%%%%%%%%%%%%%
\begin{figure}

\includegraphics[]{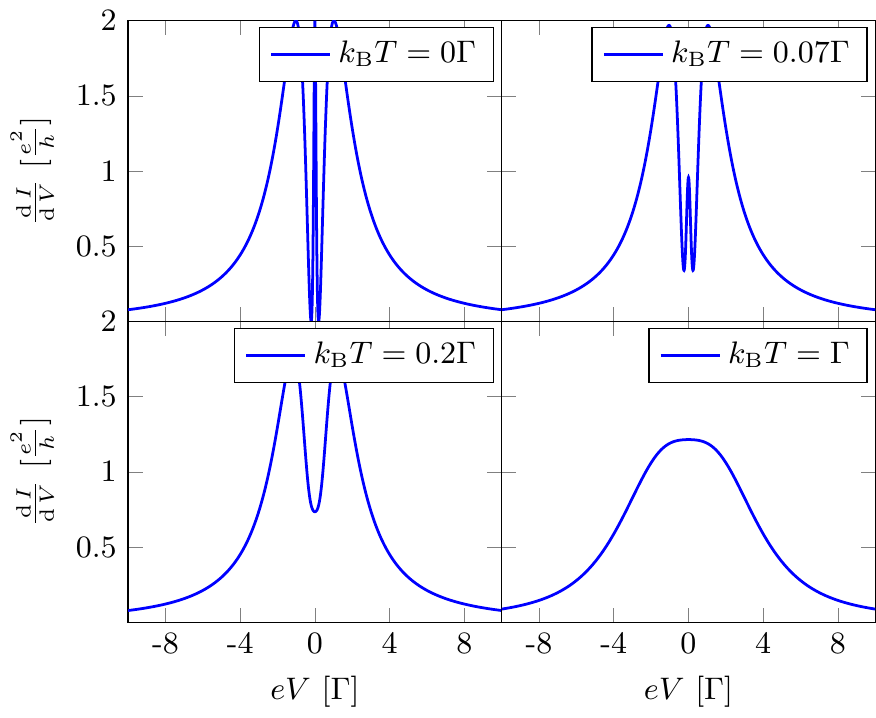}
 \caption{Differential conductance vs. bias voltage for different temperatures and $\varepsilon_\text{D}=0\Gamma$. The other parameters for these plots are $\varepsilon=0.5\Gamma$ and $|t_2|=0.1\Gamma$. }
 \label{ZBP}
\end{figure}
%%%%%%%%%%%%%%%%%%%%%%%%%%%%%%%%%%%%%%%%%%%%%%%%%%%%%%%%%%%%%%%%
with $\mathcal{B}(t)=\exp(2A t)$ we can calculate the unperturbed Majorana Green's function $D^{(0)}(t)$. For the calculation of transport properties only energy space properties are used such that we need to consider the Fourier
transform of the Majorana Green's function. The inverse of the unperturbed Majorana Green's function in Keldysh matrix representation is given by
\begin{equation}
 [D^{(0)}]^{-1}(\omega)=\left(\begin{matrix}
                               [D^{(0)}]^{--}(\omega)&0\\
                               0&-[D^{(0)}]^{--}(\omega)
                              \end{matrix}\right),
\end{equation}
where $[D^{(0)}]^{--}(\omega)=i A+\frac{\omega}{2}\mathbb{I}$.\\
The self energy containing the counting fields is given by
\begin{align}
 \Sigma^{\vec\lambda}_{\alpha\beta}(t,t')=\sum_{\delta}&-t_{\delta\alpha} {t_{\delta\beta}}^*{e}^{-i\frac{\lambda_\delta(t)-\lambda_\delta(t')}{2}}G^{(0)}_\delta(t,t')\notag\\
 &+t_{\delta\beta} {t_{\delta\alpha}}^*e^{i\frac{\lambda_\delta(t)-\lambda_\delta(t')}{2}}G^{(0)}_\delta(t',t),
\end{align}
where $G^{(0)}_\delta(t',t)$ is the unperturbed leads Green's function. Its Fourier transform therefore reads
\begin{align}
  \Sigma^\lambda(\omega)_{\mu\nu}=\sum_\alpha&\left(\begin{smallmatrix}
                         -t_{\alpha\mu} t_{\alpha\nu}^*G_\alpha^{--}(-\omega)&+t_{\alpha\nu} t_{\alpha\mu}^*e^{i\lambda_\alpha}G_\alpha^{+-}(-\omega)\\
                         t_{\alpha\nu} t_{\alpha\nu}^*{e}^{-i\lambda_\alpha}G_\alpha^{-+}(-\omega)&  -t_{\alpha\mu} t_{\alpha\nu}^*G_\alpha^{++}(-\omega)
                        \end{smallmatrix}\right)\notag\\
                        &+\left(\begin{smallmatrix}
                         t_{\alpha\nu} t_{\alpha\mu}^*G_\alpha^{--}(\omega)&-t_{\alpha\mu} t_{\alpha\nu}^*{e}^{-i\lambda_\alpha}G_\alpha^{-+}(\omega)\\
                         -t_{\alpha\nu} t_{\alpha\mu}^*e^{i\lambda_\alpha}G_\alpha^{+-}(\omega)&  +t_{\alpha\nu} t_{\alpha\mu}^*G_\alpha^{++}(\omega)
                        \end{smallmatrix}\right).
\end{align}
For a constant density of states $\rho_0=1/(2\pi v_f)$ inside the leads, the leads Green's function are
\begin{align}
 G_\alpha(\omega)=i2\pi\rho_0\left(\begin{smallmatrix}n_\alpha(\omega)-\frac{1}{2}&n_\alpha(\omega)\\
						     n_\alpha(\omega)-1&n_\alpha(\omega)-\frac{1}{2}\end{smallmatrix}\right).
\end{align}

If we now enter this into Eqs.~(\ref{Dyson}) and~(\ref{levitov}) and calculate the determinant we obtain Eq.~(\ref{cgfcalc}). Another formalism to obtain the current and noise
in the normal conducting lead - topological superconductor junction using Keldysh Green's functions is given in Ref.~\cite{PhysRevB.94.014502}.
%%%%%%%%%%%%%%%%%%%%%%%%%%%%%%%%%%%%%%%%%%%%%%%%%%%%
\section{B. Zero Bias Peak (ZBP)}
\label{secZBP}
For a finite Majorana overlap $\varepsilon\neq0$, the conductance at zero bias and temperature vanishes as long as the dot level energy is finite \cite{Pikulin2012}. However, in the case of
$\varepsilon_\text{D}=0$ the differential conductance is quantized to $2e^2/h$ for finite $|t_2|$. At finite temperature, the situation is different as the temperature reduces the height of the resonances in the
differential conductance. It is important to note that the resonance peak corresponding to $\varepsilon_\text{D}=0$ is reduced much faster than the resonances corresponding to the MBS (see 
Fig.~\ref{ZBP}). And for a temperature higher than the Majorana overlap energy ($k_{\rm B}T>\varepsilon$) the two peaks are no longer resolved and appear as single ZBP.


\begin{thebibliography}{49}%
\makeatletter
\providecommand \@ifxundefined [1]{%
 \@ifx{#1\undefined}
}%
\providecommand \@ifnum [1]{%
 \ifnum #1\expandafter \@firstoftwo
 \else \expandafter \@secondoftwo
 \fi
}%
\providecommand \@ifx [1]{%
 \ifx #1\expandafter \@firstoftwo
 \else \expandafter \@secondoftwo
 \fi
}%
\providecommand \natexlab [1]{#1}%
\providecommand \enquote  [1]{``#1''}%
\providecommand \bibnamefont  [1]{#1}%
\providecommand \bibfnamefont [1]{#1}%
\providecommand \citenamefont [1]{#1}%
\providecommand \href@noop [0]{\@secondoftwo}%
\providecommand \href [0]{\begingroup \@sanitize@url \@href}%
\providecommand \@href[1]{\@@startlink{#1}\@@href}%
\providecommand \@@href[1]{\endgroup#1\@@endlink}%
\providecommand \@sanitize@url [0]{\catcode `\\12\catcode `\$12\catcode
  `\&12\catcode `\#12\catcode `\^12\catcode `\_12\catcode `\%12\relax}%
\providecommand \@@startlink[1]{}%
\providecommand \@@endlink[0]{}%
\providecommand \url  [0]{\begingroup\@sanitize@url \@url }%
\providecommand \@url [1]{\endgroup\@href {#1}{\urlprefix }}%
\providecommand \urlprefix  [0]{URL }%
\providecommand \Eprint [0]{\href }%
\providecommand \doibase [0]{http://dx.doi.org/}%
\providecommand \selectlanguage [0]{\@gobble}%
\providecommand \bibinfo  [0]{\@secondoftwo}%
\providecommand \bibfield  [0]{\@secondoftwo}%
\providecommand \translation [1]{[#1]}%
\providecommand \BibitemOpen [0]{}%
\providecommand \bibitemStop [0]{}%
\providecommand \bibitemNoStop [0]{.\EOS\space}%
\providecommand \EOS [0]{\spacefactor3000\relax}%
\providecommand \BibitemShut  [1]{\csname bibitem#1\endcsname}%
\let\auto@bib@innerbib\@empty
%</preamble>
\bibitem [{\citenamefont {Majorana}(1937)}]{Majorana1937}%
  \BibitemOpen
  \bibfield  {author} {\bibinfo {author} {\bibfnamefont {E.}~\bibnamefont
  {Majorana}},\ }\href {\doibase 10.1007/bf02961314} {\bibfield  {journal}
  {\bibinfo  {journal} {Nuovo Cimento}\ }\textbf {\bibinfo {volume} {14}},\
  \bibinfo {pages} {171} (\bibinfo {year} {1937})}\BibitemShut {NoStop}%
\bibitem [{\citenamefont {Nayak}\ \emph {et~al.}(2008)\citenamefont {Nayak},
  \citenamefont {Simon}, \citenamefont {Stern}, \citenamefont {Freedman},\ and\
  \citenamefont {Das~Sarma}}]{Nayak2008}%
  \BibitemOpen
  \bibfield  {author} {\bibinfo {author} {\bibfnamefont {C.}~\bibnamefont
  {Nayak}}, \bibinfo {author} {\bibfnamefont {S.~H.}\ \bibnamefont {Simon}},
  \bibinfo {author} {\bibfnamefont {A.}~\bibnamefont {Stern}}, \bibinfo
  {author} {\bibfnamefont {M.}~\bibnamefont {Freedman}}, \ and\ \bibinfo
  {author} {\bibfnamefont {S.}~\bibnamefont {Das~Sarma}},\ }\href {\doibase
  10.1103/revmodphys.80.1083} {\bibfield  {journal} {\bibinfo  {journal} {Rev.
  Mod. Phys.}\ }\textbf {\bibinfo {volume} {80}},\ \bibinfo {pages} {1083}
  (\bibinfo {year} {2008})}\BibitemShut {NoStop}%
\bibitem [{\citenamefont {Pachos}(2013)}]{Pachos2012}%
  \BibitemOpen
  \bibfield  {author} {\bibinfo {author} {\bibfnamefont {J.~K.}\ \bibnamefont
  {Pachos}},\ }\enquote {\bibinfo {title} {Introduction to topological quantum
  computation},}\ \ (\bibinfo  {publisher} {Cambridge University Press},\
  \bibinfo {year} {2013})\BibitemShut {NoStop}%
\bibitem [{\citenamefont {Mourik}\ \emph {et~al.}(2012)\citenamefont {Mourik},
  \citenamefont {Zuo}, \citenamefont {Frolov}, \citenamefont {Plissard},
  \citenamefont {Bakkers},\ and\ \citenamefont {Kouwenhoven}}]{Mourik2012}%
  \BibitemOpen
  \bibfield  {author} {\bibinfo {author} {\bibfnamefont {V.}~\bibnamefont
  {Mourik}}, \bibinfo {author} {\bibfnamefont {K.}~\bibnamefont {Zuo}},
  \bibinfo {author} {\bibfnamefont {S.~M.}\ \bibnamefont {Frolov}}, \bibinfo
  {author} {\bibfnamefont {S.~R.}\ \bibnamefont {Plissard}}, \bibinfo {author}
  {\bibfnamefont {E.~P. A.~M.}\ \bibnamefont {Bakkers}}, \ and\ \bibinfo
  {author} {\bibfnamefont {L.~P.}\ \bibnamefont {Kouwenhoven}},\ }\href
  {\doibase 10.1126/science.1222360} {\bibfield  {journal} {\bibinfo  {journal}
  {Science}\ }\textbf {\bibinfo {volume} {336}},\ \bibinfo {pages} {1003}
  (\bibinfo {year} {2012})}\BibitemShut {NoStop}%
\bibitem [{\citenamefont {Das}\ \emph {et~al.}(2012)\citenamefont {Das},
  \citenamefont {Ronen}, \citenamefont {Most}, \citenamefont {Oreg},
  \citenamefont {Heiblum},\ and\ \citenamefont {Shtrikman}}]{Das2012}%
  \BibitemOpen
  \bibfield  {author} {\bibinfo {author} {\bibfnamefont {A.}~\bibnamefont
  {Das}}, \bibinfo {author} {\bibfnamefont {Y.}~\bibnamefont {Ronen}}, \bibinfo
  {author} {\bibfnamefont {Y.}~\bibnamefont {Most}}, \bibinfo {author}
  {\bibfnamefont {Y.}~\bibnamefont {Oreg}}, \bibinfo {author} {\bibfnamefont
  {M.}~\bibnamefont {Heiblum}}, \ and\ \bibinfo {author} {\bibfnamefont
  {H.}~\bibnamefont {Shtrikman}},\ }\href {\doibase 10.1038/nphys2479}
  {\bibfield  {journal} {\bibinfo  {journal} {Nat. Phys.}\ }\textbf {\bibinfo
  {volume} {8}},\ \bibinfo {pages} {887–895} (\bibinfo {year}
  {2012})}\BibitemShut {NoStop}%
\bibitem [{\citenamefont {Lee}\ \emph {et~al.}(2013)\citenamefont {Lee},
  \citenamefont {Jiang}, \citenamefont {Houzet}, \citenamefont {Aguado},
  \citenamefont {Lieber},\ and\ \citenamefont {De~Franceschi}}]{Lee2013}%
  \BibitemOpen
  \bibfield  {author} {\bibinfo {author} {\bibfnamefont {E.~J.~H.}\
  \bibnamefont {Lee}}, \bibinfo {author} {\bibfnamefont {X.}~\bibnamefont
  {Jiang}}, \bibinfo {author} {\bibfnamefont {M.}~\bibnamefont {Houzet}},
  \bibinfo {author} {\bibfnamefont {R.}~\bibnamefont {Aguado}}, \bibinfo
  {author} {\bibfnamefont {C.~M.}\ \bibnamefont {Lieber}}, \ and\ \bibinfo
  {author} {\bibfnamefont {S.}~\bibnamefont {De~Franceschi}},\ }\href {\doibase
  10.1038/nnano.2013.267} {\bibfield  {journal} {\bibinfo  {journal} {Nat.
  Nanotech.}\ }\textbf {\bibinfo {volume} {9}},\ \bibinfo {pages} {79–84}
  (\bibinfo {year} {2013})}\BibitemShut {NoStop}%
\bibitem [{\citenamefont {Nadj-Perge}\ \emph {et~al.}(2014)\citenamefont
  {Nadj-Perge}, \citenamefont {Drozdov}, \citenamefont {Li}, \citenamefont
  {Chen}, \citenamefont {Jeon}, \citenamefont {Seo}, \citenamefont {MacDonald},
  \citenamefont {Bernevig},\ and\ \citenamefont {Yazdani}}]{Nadj-Perge2014}%
  \BibitemOpen
  \bibfield  {author} {\bibinfo {author} {\bibfnamefont {S.}~\bibnamefont
  {Nadj-Perge}}, \bibinfo {author} {\bibfnamefont {I.~K.}\ \bibnamefont
  {Drozdov}}, \bibinfo {author} {\bibfnamefont {J.}~\bibnamefont {Li}},
  \bibinfo {author} {\bibfnamefont {H.}~\bibnamefont {Chen}}, \bibinfo {author}
  {\bibfnamefont {S.}~\bibnamefont {Jeon}}, \bibinfo {author} {\bibfnamefont
  {J.}~\bibnamefont {Seo}}, \bibinfo {author} {\bibfnamefont {A.~H.}\
  \bibnamefont {MacDonald}}, \bibinfo {author} {\bibfnamefont {B.~A.}\
  \bibnamefont {Bernevig}}, \ and\ \bibinfo {author} {\bibfnamefont
  {A.}~\bibnamefont {Yazdani}},\ }\href@noop {} {\bibfield  {journal} {\bibinfo
   {journal} {Science}\ }\textbf {\bibinfo {volume} {346}},\ \bibinfo {pages}
  {602} (\bibinfo {year} {2014})}\BibitemShut {NoStop}%
\bibitem [{\citenamefont {{Pawlak}}\ \emph {et~al.}(2016)\citenamefont
  {{Pawlak}}, \citenamefont {{Kisiel}}, \citenamefont {{Klinovaja}},
  \citenamefont {{Meier}}, \citenamefont {{Kawai}}, \citenamefont {{Glatzel}},
  \citenamefont {{Loss}},\ and\ \citenamefont {{Meyer}}}]{Pawlak2015}%
  \BibitemOpen
  \bibfield  {author} {\bibinfo {author} {\bibfnamefont {R.}~\bibnamefont
  {{Pawlak}}}, \bibinfo {author} {\bibfnamefont {M.}~\bibnamefont {{Kisiel}}},
  \bibinfo {author} {\bibfnamefont {J.}~\bibnamefont {{Klinovaja}}}, \bibinfo
  {author} {\bibfnamefont {T.}~\bibnamefont {{Meier}}}, \bibinfo {author}
  {\bibfnamefont {S.}~\bibnamefont {{Kawai}}}, \bibinfo {author} {\bibfnamefont
  {T.}~\bibnamefont {{Glatzel}}}, \bibinfo {author} {\bibfnamefont
  {D.}~\bibnamefont {{Loss}}}, \ and\ \bibinfo {author} {\bibfnamefont
  {E.}~\bibnamefont {{Meyer}}},\ }\href@noop {} {\bibfield  {journal} {\bibinfo
   {journal} {npj Quantum Information}\ }\textbf {\bibinfo {volume} {2}},\
  \bibinfo {pages} {16035} (\bibinfo {year} {2016})}\BibitemShut {NoStop}%
\bibitem [{\citenamefont {Kitaev}(2001)}]{Kitaev2001}%
  \BibitemOpen
  \bibfield  {author} {\bibinfo {author} {\bibfnamefont {A.~Y.}\ \bibnamefont
  {Kitaev}},\ }\href {\doibase 10.1070/1063-7869/44/10s/s29} {\bibfield
  {journal} {\bibinfo  {journal} {Phys. Usp.}\ }\textbf {\bibinfo {volume}
  {44}},\ \bibinfo {pages} {131} (\bibinfo {year} {2001})}\BibitemShut
  {NoStop}%
\bibitem [{\citenamefont {Fu}\ and\ \citenamefont {Kane}(2009)}]{Fu2009b}%
  \BibitemOpen
  \bibfield  {author} {\bibinfo {author} {\bibfnamefont {L.}~\bibnamefont
  {Fu}}\ and\ \bibinfo {author} {\bibfnamefont {C.~L.}\ \bibnamefont {Kane}},\
  }\href {\doibase 10.1103/PhysRevB.79.161408} {\bibfield  {journal} {\bibinfo
  {journal} {Phys. Rev. B}\ }\textbf {\bibinfo {volume} {79}},\ \bibinfo
  {pages} {161408} (\bibinfo {year} {2009})}\BibitemShut {NoStop}%
\bibitem [{\citenamefont {Beenakker}\ \emph {et~al.}(2013)\citenamefont
  {Beenakker}, \citenamefont {Pikulin}, \citenamefont {Hyart}, \citenamefont
  {Schomerus},\ and\ \citenamefont {Dahlhaus}}]{PhysRevLett.110.017003}%
  \BibitemOpen
  \bibfield  {author} {\bibinfo {author} {\bibfnamefont {C.~W.~J.}\
  \bibnamefont {Beenakker}}, \bibinfo {author} {\bibfnamefont {D.~I.}\
  \bibnamefont {Pikulin}}, \bibinfo {author} {\bibfnamefont {T.}~\bibnamefont
  {Hyart}}, \bibinfo {author} {\bibfnamefont {H.}~\bibnamefont {Schomerus}}, \
  and\ \bibinfo {author} {\bibfnamefont {J.~P.}\ \bibnamefont {Dahlhaus}},\
  }\href {\doibase 10.1103/PhysRevLett.110.017003} {\bibfield  {journal}
  {\bibinfo  {journal} {Phys. Rev. Lett.}\ }\textbf {\bibinfo {volume} {110}},\
  \bibinfo {pages} {017003} (\bibinfo {year} {2013})}\BibitemShut {NoStop}%
\bibitem [{\citenamefont {Cr\'epin}\ and\ \citenamefont
  {Trauzettel}(2014)}]{PhysRevLett.112.077002}%
  \BibitemOpen
  \bibfield  {author} {\bibinfo {author} {\bibfnamefont {F.}\
  \bibnamefont {Cr\'epin}}\ and\ \bibinfo {author} {\bibfnamefont
  {B.}~\bibnamefont {Trauzettel}},\ }\href {\doibase
  10.1103/PhysRevLett.112.077002} {\bibfield  {journal} {\bibinfo  {journal}
  {Phys. Rev. Lett.}\ }\textbf {\bibinfo {volume} {112}},\ \bibinfo {pages}
  {077002} (\bibinfo {year} {2014})}\BibitemShut {NoStop}%
\bibitem [{\citenamefont {Rokhinson}\ \emph {et~al.}(2012)\citenamefont
  {Rokhinson}, \citenamefont {Liu},\ and\ \citenamefont
  {Furdyna}}]{Rokhinson2012}%
  \BibitemOpen
  \bibfield  {author} {\bibinfo {author} {\bibfnamefont {L.~P.}\ \bibnamefont
  {Rokhinson}}, \bibinfo {author} {\bibfnamefont {X.}~\bibnamefont {Liu}}, \
  and\ \bibinfo {author} {\bibfnamefont {J.~K.}\ \bibnamefont {Furdyna}},\
  }\href {\doibase 10.1038/nphys2429} {\bibfield  {journal} {\bibinfo
  {journal} {Nat. Phys.}\ }\textbf {\bibinfo {volume} {8}},\ \bibinfo {pages}
  {795–799} (\bibinfo {year} {2012})}\BibitemShut {NoStop}%
\bibitem [{\citenamefont {Wiedenmann}\ \emph {et~al.}(2016)\citenamefont
  {Wiedenmann}, \citenamefont {Bocquillon}, \citenamefont {Deacon},
  \citenamefont {Hartinger}, \citenamefont {Herrmann}, \citenamefont
  {Klapwijk}, \citenamefont {Maier}, \citenamefont {Ames}, \citenamefont
  {Br\"une}, \citenamefont {Gould}, \citenamefont {Oiwa}, \citenamefont
  {Ishibashi}, \citenamefont {Tarucha}, \citenamefont {Buhmann},\ and\
  \citenamefont {Molenkamp}}]{Wiedenmann2016}%
  \BibitemOpen
  \bibfield  {author} {\bibinfo {author} {\bibfnamefont {J.}~\bibnamefont
  {Wiedenmann}}, \bibinfo {author} {\bibfnamefont {E.}~\bibnamefont
  {Bocquillon}}, \bibinfo {author} {\bibfnamefont {R.~S.}\ \bibnamefont
  {Deacon}}, \bibinfo {author} {\bibfnamefont {S.}~\bibnamefont {Hartinger}},
  \bibinfo {author} {\bibfnamefont {O.}~\bibnamefont {Herrmann}}, \bibinfo
  {author} {\bibfnamefont {T.~M.}\ \bibnamefont {Klapwijk}}, \bibinfo {author}
  {\bibfnamefont {L.}~\bibnamefont {Maier}}, \bibinfo {author} {\bibfnamefont
  {C.}~\bibnamefont {Ames}}, \bibinfo {author} {\bibfnamefont {C.}~\bibnamefont
  {Br\"une}}, \bibinfo {author} {\bibfnamefont {C.}~\bibnamefont {Gould}},
  \bibinfo {author} {\bibfnamefont {A.}~\bibnamefont {Oiwa}}, \bibinfo {author}
  {\bibfnamefont {K.}~\bibnamefont {Ishibashi}}, \bibinfo {author}
  {\bibfnamefont {S.}~\bibnamefont {Tarucha}}, \bibinfo {author} {\bibfnamefont
  {H.}~\bibnamefont {Buhmann}}, \ and\ \bibinfo {author} {\bibfnamefont
  {L.~W.}\ \bibnamefont {Molenkamp}},\ }\href {\doibase 10.1038/ncomms10303}
  {\bibfield  {journal} {\bibinfo  {journal} {Nat. Comun.}\ }\textbf {\bibinfo
  {volume} {7}},\ \bibinfo {pages} {10303} (\bibinfo {year}
  {2016})}\BibitemShut {NoStop}%
\bibitem [{\citenamefont {Das~Sarma}\ \emph {et~al.}(2005)\citenamefont
  {Das~Sarma}, \citenamefont {Freedman},\ and\ \citenamefont
  {Nayak}}]{DasSarma2005}%
  \BibitemOpen
  \bibfield  {author} {\bibinfo {author} {\bibfnamefont {S.}~\bibnamefont
  {Das~Sarma}}, \bibinfo {author} {\bibfnamefont {M.}~\bibnamefont {Freedman}},
  \ and\ \bibinfo {author} {\bibfnamefont {C.}~\bibnamefont {Nayak}},\ }\href
  {\doibase 10.1103/PhysRevLett.94.166802} {\bibfield  {journal} {\bibinfo
  {journal} {Phys. Rev. Lett.}\ }\textbf {\bibinfo {volume} {94}},\ \bibinfo
  {pages} {166802} (\bibinfo {year} {2005})}\BibitemShut {NoStop}%
\bibitem [{\citenamefont {Stern}\ and\ \citenamefont
  {Halperin}(2006)}]{Stern2006}%
  \BibitemOpen
  \bibfield  {author} {\bibinfo {author} {\bibfnamefont {A.}~\bibnamefont
  {Stern}}\ and\ \bibinfo {author} {\bibfnamefont {B.~I.}\ \bibnamefont
  {Halperin}},\ }\href {\doibase 10.1103/PhysRevLett.96.016802} {\bibfield
  {journal} {\bibinfo  {journal} {Phys. Rev. Lett.}\ }\textbf {\bibinfo
  {volume} {96}},\ \bibinfo {pages} {016802} (\bibinfo {year}
  {2006})}\BibitemShut {NoStop}%
\bibitem [{\citenamefont {Bonderson}\ \emph {et~al.}(2006)\citenamefont
  {Bonderson}, \citenamefont {Kitaev},\ and\ \citenamefont
  {Shtengel}}]{Bonderson2006}%
  \BibitemOpen
  \bibfield  {author} {\bibinfo {author} {\bibfnamefont {P.}~\bibnamefont
  {Bonderson}}, \bibinfo {author} {\bibfnamefont {A.}~\bibnamefont {Kitaev}}, \
  and\ \bibinfo {author} {\bibfnamefont {K.}~\bibnamefont {Shtengel}},\ }\href
  {\doibase 10.1103/PhysRevLett.96.016803} {\bibfield  {journal} {\bibinfo
  {journal} {Phys. Rev. Lett.}\ }\textbf {\bibinfo {volume} {96}},\ \bibinfo
  {pages} {016803} (\bibinfo {year} {2006})}\BibitemShut {NoStop}%
\bibitem [{\citenamefont {Alicea}\ \emph {et~al.}(2011)\citenamefont {Alicea},
  \citenamefont {Oreg}, \citenamefont {Refael}, \citenamefont {von Oppen},\
  and\ \citenamefont {Fisher}}]{Alicea2011}%
  \BibitemOpen
  \bibfield  {author} {\bibinfo {author} {\bibfnamefont {J.}~\bibnamefont
  {Alicea}}, \bibinfo {author} {\bibfnamefont {Y.}~\bibnamefont {Oreg}},
  \bibinfo {author} {\bibfnamefont {G.}~\bibnamefont {Refael}}, \bibinfo
  {author} {\bibfnamefont {F.}~\bibnamefont {von Oppen}}, \ and\ \bibinfo
  {author} {\bibfnamefont {M.~P.~A.}\ \bibnamefont {Fisher}},\ }\href {\doibase
  10.1038/nphys1915} {\bibfield  {journal} {\bibinfo  {journal} {Nat. Phys.}\
  }\textbf {\bibinfo {volume} {7}},\ \bibinfo {pages} {412} (\bibinfo {year}
  {2011})}\BibitemShut {NoStop}%
\bibitem [{\citenamefont {Sau}\ \emph {et~al.}(2011)\citenamefont {Sau},
  \citenamefont {Clarke},\ and\ \citenamefont {Tewari}}]{Sau2011}%
  \BibitemOpen
  \bibfield  {author} {\bibinfo {author} {\bibfnamefont {J.~D.}\ \bibnamefont
  {Sau}}, \bibinfo {author} {\bibfnamefont {D.~J.}\ \bibnamefont {Clarke}}, \
  and\ \bibinfo {author} {\bibfnamefont {S.}~\bibnamefont {Tewari}},\ }\href
  {\doibase 10.1103/PhysRevB.84.094505} {\bibfield  {journal} {\bibinfo
  {journal} {Phys. Rev. B}\ }\textbf {\bibinfo {volume} {84}},\ \bibinfo
  {pages} {094505} (\bibinfo {year} {2011})}\BibitemShut {NoStop}%
\bibitem [{\citenamefont {van Heck}\ \emph {et~al.}(2012)\citenamefont {van
  Heck}, \citenamefont {Akhmerov}, \citenamefont {Hassler}, \citenamefont
  {Burrello},\ and\ \citenamefont {Beenakker}}]{vanHeck2012}%
  \BibitemOpen
  \bibfield  {author} {\bibinfo {author} {\bibfnamefont {B.}~\bibnamefont {van
  Heck}}, \bibinfo {author} {\bibfnamefont {A.~R.}\ \bibnamefont {Akhmerov}},
  \bibinfo {author} {\bibfnamefont {F.}~\bibnamefont {Hassler}}, \bibinfo
  {author} {\bibfnamefont {M.}~\bibnamefont {Burrello}}, \ and\ \bibinfo
  {author} {\bibfnamefont {C.~W.~J.}\ \bibnamefont {Beenakker}},\ }\href
  {http://stacks.iop.org/1367-2630/14/i=3/a=035019} {\bibfield  {journal}
  {\bibinfo  {journal} {New J. Phys.}\ }\textbf {\bibinfo {volume} {14}},\
  \bibinfo {pages} {035019} (\bibinfo {year} {2012})}\BibitemShut {NoStop}%
\bibitem [{\citenamefont {Cheng}\ \emph {et~al.}(2012)\citenamefont {Cheng},
  \citenamefont {Lutchyn},\ and\ \citenamefont {Das~Sarma}}]{Cheng2012}%
  \BibitemOpen
  \bibfield  {author} {\bibinfo {author} {\bibfnamefont {M.}~\bibnamefont
  {Cheng}}, \bibinfo {author} {\bibfnamefont {R.~M.}\ \bibnamefont {Lutchyn}},
  \ and\ \bibinfo {author} {\bibfnamefont {S.}~\bibnamefont {Das~Sarma}},\
  }\href {\doibase 10.1103/PhysRevB.85.165124} {\bibfield  {journal} {\bibinfo
  {journal} {Phys. Rev. B}\ }\textbf {\bibinfo {volume} {85}},\ \bibinfo
  {pages} {165124} (\bibinfo {year} {2012})}\BibitemShut {NoStop}%
\bibitem [{\citenamefont {Pedrocchi}\ and\ \citenamefont
  {DiVincenzo}(2015)}]{Pedrocchi2015}%
  \BibitemOpen
  \bibfield  {author} {\bibinfo {author} {\bibfnamefont {F.~L.}\ \bibnamefont
  {Pedrocchi}}\ and\ \bibinfo {author} {\bibfnamefont {D.~P.}\ \bibnamefont
  {DiVincenzo}},\ }\href {\doibase 10.1103/PhysRevLett.115.120402} {\bibfield
  {journal} {\bibinfo  {journal} {Phys. Rev. Lett.}\ }\textbf {\bibinfo
  {volume} {115}},\ \bibinfo {pages} {120402} (\bibinfo {year}
  {2015})}\BibitemShut {NoStop}%
\bibitem [{\citenamefont {Lutchyn}\ \emph {et~al.}(2010)\citenamefont
  {Lutchyn}, \citenamefont {Sau},\ and\ \citenamefont
  {Das~Sarma}}]{Lutchyn2010}%
  \BibitemOpen
  \bibfield  {author} {\bibinfo {author} {\bibfnamefont {R.~M.}\ \bibnamefont
  {Lutchyn}}, \bibinfo {author} {\bibfnamefont {J.~D.}\ \bibnamefont {Sau}}, \
  and\ \bibinfo {author} {\bibfnamefont {S.}~\bibnamefont {Das~Sarma}},\
  }\href@noop {} {\bibfield  {journal} {\bibinfo  {journal} {Phys. Rev. Lett.}\
  }\textbf {\bibinfo {volume} {105}},\ \bibinfo {pages} {077001}  (\bibinfo {year} {2010})}\BibitemShut
  {NoStop}%
\bibitem [{\citenamefont {Oreg}\ \emph {et~al.}(2010)\citenamefont {Oreg},
  \citenamefont {Refael},\ and\ \citenamefont {von Oppen}}]{Oreg2010}%
  \BibitemOpen
  \bibfield  {author} {\bibinfo {author} {\bibfnamefont {Y.}~\bibnamefont
  {Oreg}}, \bibinfo {author} {\bibfnamefont {G.}~\bibnamefont {Refael}}, \ and\
  \bibinfo {author} {\bibfnamefont {F.}~\bibnamefont {von Oppen}},\ }\href
  {\doibase 10.1103/physrevlett.105.177002} {\bibfield  {journal} {\bibinfo
  {journal} {Phys. Rev. Lett}\ }\textbf {\bibinfo {volume} {105}},\ \bibinfo
  {pages} {177002} (\bibinfo {year} {2010})}\BibitemShut {NoStop}%
\bibitem [{\citenamefont {Brouwer}\ \emph {et~al.}(2011)\citenamefont
  {Brouwer}, \citenamefont {Duckheim}, \citenamefont {Romito},\ and\
  \citenamefont {von Oppen}}]{PhysRevLett.107.196804}%
  \BibitemOpen
  \bibfield  {author} {\bibinfo {author} {\bibfnamefont {P.~W.}\ \bibnamefont
  {Brouwer}}, \bibinfo {author} {\bibfnamefont {M.}~\bibnamefont {Duckheim}},
  \bibinfo {author} {\bibfnamefont {A.}~\bibnamefont {Romito}}, \ and\ \bibinfo
  {author} {\bibfnamefont {F.}~\bibnamefont {von Oppen}},\ }\href {\doibase
  10.1103/PhysRevLett.107.196804} {\bibfield  {journal} {\bibinfo  {journal}
  {Phys. Rev. Lett.}\ }\textbf {\bibinfo {volume} {107}},\ \bibinfo {pages}
  {196804} (\bibinfo {year} {2011})}\BibitemShut {NoStop}%
\bibitem [{\citenamefont {Rainis}\ \emph {et~al.}(2013)\citenamefont {Rainis},
  \citenamefont {Trifunovic}, \citenamefont {Klinovaja},\ and\ \citenamefont
  {Loss}}]{PhysRevB.87.024515}%
  \BibitemOpen
  \bibfield  {author} {\bibinfo {author} {\bibfnamefont {D.}~\bibnamefont
  {Rainis}}, \bibinfo {author} {\bibfnamefont {L.}~\bibnamefont {Trifunovic}},
  \bibinfo {author} {\bibfnamefont {J.}~\bibnamefont {Klinovaja}}, \ and\
  \bibinfo {author} {\bibfnamefont {D.}~\bibnamefont {Loss}},\ }\href {\doibase
  10.1103/PhysRevB.87.024515} {\bibfield  {journal} {\bibinfo  {journal} {Phys.
  Rev. B}\ }\textbf {\bibinfo {volume} {87}},\ \bibinfo {pages} {024515}
  (\bibinfo {year} {2013})}\BibitemShut {NoStop}%
\bibitem [{\citenamefont {Nilsson}\ \emph {et~al.}(2008)\citenamefont
  {Nilsson}, \citenamefont {Akhmerov},\ and\ \citenamefont
  {Beenakker}}]{Nilsson2008}%
  \BibitemOpen
  \bibfield  {author} {\bibinfo {author} {\bibfnamefont {J.}~\bibnamefont
  {Nilsson}}, \bibinfo {author} {\bibfnamefont {A.}~\bibnamefont {Akhmerov}}, \
  and\ \bibinfo {author} {\bibfnamefont {C.}~\bibnamefont {Beenakker}},\ }\href
  {\doibase 10.1103/physrevlett.101.120403} {\bibfield  {journal} {\bibinfo
  {journal} {Phys. Rev. Lett}\ }\textbf {\bibinfo {volume} {101}},\ \bibinfo
  {pages} {120403} (\bibinfo {year} {2008})}\BibitemShut {NoStop}%
\bibitem [{\citenamefont {Fu}(2010)}]{Fu2010}%
  \BibitemOpen
  \bibfield  {author} {\bibinfo {author} {\bibfnamefont {L.}~\bibnamefont
  {Fu}},\ }\href {http://dx.doi.org/10.1103/PhysRevLett.104.056402} {\bibfield
  {journal} {\bibinfo  {journal} {Phys. Rev. Let.}\ }\textbf {\bibinfo {volume}
  {104}},\ \bibinfo {pages} {056402} (\bibinfo {year} {2010})}\BibitemShut
  {NoStop}%
\bibitem [{\citenamefont {Albrecht}\ \emph {et~al.}(2016)\citenamefont
  {Albrecht}, \citenamefont {Higginbotham}, \citenamefont {Madsen},
  \citenamefont {Kuemmeth}, \citenamefont {Jespersen}, \citenamefont
  {Nyg\r{a}rd}, \citenamefont {Krogstrup},\ and\ \citenamefont
  {Marcus}}]{Albrecht2016}%
  \BibitemOpen
  \bibfield  {author} {\bibinfo {author} {\bibfnamefont {S.~M.}\ \bibnamefont
  {Albrecht}}, \bibinfo {author} {\bibfnamefont {A.~P.}\ \bibnamefont
  {Higginbotham}}, \bibinfo {author} {\bibfnamefont {M.}~\bibnamefont
  {Madsen}}, \bibinfo {author} {\bibfnamefont {F.}~\bibnamefont {Kuemmeth}},
  \bibinfo {author} {\bibfnamefont {T.~S.}\ \bibnamefont {Jespersen}}, \bibinfo
  {author} {\bibfnamefont {J.}~\bibnamefont {Nyg\r{a}rd}}, \bibinfo {author}
  {\bibfnamefont {P.}~\bibnamefont {Krogstrup}}, \ and\ \bibinfo {author}
  {\bibfnamefont {C.~M.}\ \bibnamefont {Marcus}},\ }\href {\doibase
  10.1038/nature17162} {\bibfield  {journal} {\bibinfo  {journal} {Nature}\
  }\textbf {\bibinfo {volume} {531}},\ \bibinfo {pages} {206–209} (\bibinfo
  {year} {2016})}\BibitemShut {NoStop}%
\bibitem [{\citenamefont {Park}\ and\ \citenamefont {Recher}(2015)}]{Park2015}%
  \BibitemOpen
  \bibfield  {author} {\bibinfo {author} {\bibfnamefont {S.}~\bibnamefont
  {Park}}\ and\ \bibinfo {author} {\bibfnamefont {P.}~\bibnamefont {Recher}},\
  }\href {\doibase 10.1103/PhysRevLett.115.246403} {\bibfield  {journal}
  {\bibinfo  {journal} {Phys. Rev. Lett.}\ }\textbf {\bibinfo {volume} {115}},\
  \bibinfo {pages} {246403} (\bibinfo {year} {2015})}\BibitemShut {NoStop}%
\bibitem [{\citenamefont {Liu}\ and\ \citenamefont {Baranger}(2011)}]{Liu2011}%
  \BibitemOpen
  \bibfield  {author} {\bibinfo {author} {\bibfnamefont {D.~E.}\ \bibnamefont
  {Liu}}\ and\ \bibinfo {author} {\bibfnamefont {H.~U.}\ \bibnamefont
  {Baranger}},\ }\href {http://dx.doi.org/10.1103/PhysRevB.84.201308}
  {\bibfield  {journal} {\bibinfo  {journal} {Phys. Rev. B}\ }\textbf {\bibinfo
  {volume} {84}},\ \bibinfo {pages} {201308} (\bibinfo {year}
  {2011})}\BibitemShut {NoStop}%
\bibitem [{\citenamefont {Leijnse}\ and\ \citenamefont
  {Flensberg}(2011)}]{Leijnse2011}%
  \BibitemOpen
  \bibfield  {author} {\bibinfo {author} {\bibfnamefont {M.}~\bibnamefont
  {Leijnse}}\ and\ \bibinfo {author} {\bibfnamefont {K.}~\bibnamefont
  {Flensberg}},\ }\href {http://dx.doi.org/10.1103/PhysRevB.84.140501}
  {\bibfield  {journal} {\bibinfo  {journal} {Phys. Rev. B}\ }\textbf {\bibinfo
  {volume} {84}},\ \bibinfo {pages} {140501} (\bibinfo {year}
  {2011})}\BibitemShut {NoStop}%
\bibitem [{\citenamefont {Vernek}\ \emph {et~al.}(2014)\citenamefont {Vernek},
  \citenamefont {Penteado}, \citenamefont {Seridonio},\ and\ \citenamefont
  {Egues}}]{Vernek2014}%
  \BibitemOpen
  \bibfield  {author} {\bibinfo {author} {\bibfnamefont {E.}~\bibnamefont
  {Vernek}}, \bibinfo {author} {\bibfnamefont {P.~H.}\ \bibnamefont
  {Penteado}}, \bibinfo {author} {\bibfnamefont {A.~C.}\ \bibnamefont
  {Seridonio}}, \ and\ \bibinfo {author} {\bibfnamefont {J.~C.}\ \bibnamefont
  {Egues}},\ }\href {http://dx.doi.org/10.1103/PhysRevB.89.165314} {\bibfield
  {journal} {\bibinfo  {journal} {Phys. Rev. B}\ }\textbf {\bibinfo {volume}
  {89}},\ \bibinfo {pages} {115435} (\bibinfo {year} {2014})}\BibitemShut {NoStop}%
\bibitem [{\citenamefont {Deng}\ \emph {et~al.}(2016)\citenamefont {Deng},
  \citenamefont {Vaitiekenas}, \citenamefont {Hansen}, \citenamefont {Danon},
  \citenamefont {Leijnse}, \citenamefont {Flensberg}, \citenamefont {Nyg{\r
  a}rd}, \citenamefont {Krogstrup},\ and\ \citenamefont {Marcus}}]{Deng1557}%
  \BibitemOpen
  \bibfield  {author} {\bibinfo {author} {\bibfnamefont {M.~T.}\ \bibnamefont
  {Deng}}, \bibinfo {author} {\bibfnamefont {S.}~\bibnamefont {Vaitiekenas}},
  \bibinfo {author} {\bibfnamefont {E.~B.}\ \bibnamefont {Hansen}}, \bibinfo
  {author} {\bibfnamefont {J.}~\bibnamefont {Danon}}, \bibinfo {author}
  {\bibfnamefont {M.}~\bibnamefont {Leijnse}}, \bibinfo {author} {\bibfnamefont
  {K.}~\bibnamefont {Flensberg}}, \bibinfo {author} {\bibfnamefont
  {J.}~\bibnamefont {Nyg{\r a}rd}}, \bibinfo {author} {\bibfnamefont
  {P.}~\bibnamefont {Krogstrup}}, \ and\ \bibinfo {author} {\bibfnamefont
  {C.~M.}\ \bibnamefont {Marcus}},\ }\href {\doibase 10.1126/science.aaf3961}
  {\bibfield  {journal} {\bibinfo  {journal} {Science}\ }\textbf {\bibinfo
  {volume} {354}},\ \bibinfo {pages} {1557} (\bibinfo {year}
  {2016})}\BibitemShut {NoStop}%
\bibitem [{\citenamefont {Miroshnichenko}\ \emph {et~al.}(2010)\citenamefont
  {Miroshnichenko}, \citenamefont {Flach},\ and\ \citenamefont
  {Kivshar}}]{Miroshnichenko2010}%
  \BibitemOpen
  \bibfield  {author} {\bibinfo {author} {\bibfnamefont {A.~E.}\ \bibnamefont
  {Miroshnichenko}}, \bibinfo {author} {\bibfnamefont {S.}~\bibnamefont
  {Flach}}, \ and\ \bibinfo {author} {\bibfnamefont {Y.~S.}\ \bibnamefont
  {Kivshar}},\ }\href {\doibase 10.1103/revmodphys.82.2257} {\bibfield
  {journal} {\bibinfo  {journal} {Rev. Mod. Phys.}\ }\textbf {\bibinfo {volume}
  {82}},\ \bibinfo {pages} {2257} (\bibinfo {year} {2010})}\BibitemShut
  {NoStop}%
\bibitem [{\citenamefont {Dessotti}\ \emph {et~al.}(2014)\citenamefont
  {Dessotti}, \citenamefont {Ricco}, \citenamefont {de~Souza}, \citenamefont
  {Souza},\ and\ \citenamefont {Seridonio}}]{Dessotti2014}%
  \BibitemOpen
  \bibfield  {author} {\bibinfo {author} {\bibfnamefont {F.~A.}\ \bibnamefont
  {Dessotti}}, \bibinfo {author} {\bibfnamefont {L.~S.}\ \bibnamefont {Ricco}},
  \bibinfo {author} {\bibfnamefont {M.}~\bibnamefont {de~Souza}}, \bibinfo
  {author} {\bibfnamefont {F.~M.}\ \bibnamefont {Souza}}, \ and\ \bibinfo
  {author} {\bibfnamefont {A.~C.}\ \bibnamefont {Seridonio}},\ }\href {\doibase
  10.1063/1.4898776} {\bibfield  {journal} {\bibinfo  {journal} {J. Appl.
  Phys.}\ }\textbf {\bibinfo {volume} {116}},\ \bibinfo {pages} {173701}
  (\bibinfo {year} {2014})}\BibitemShut {NoStop}%
\bibitem [{\citenamefont {Xia}\ \emph {et~al.}(2015)\citenamefont {Xia},
  \citenamefont {Duan},\ and\ \citenamefont {Zhang}}]{Xia2015}%
  \BibitemOpen
  \bibfield  {author} {\bibinfo {author} {\bibfnamefont {J.-J.}\ \bibnamefont
  {Xia}}, \bibinfo {author} {\bibfnamefont {S.-Q.}\ \bibnamefont {Duan}}, \
  and\ \bibinfo {author} {\bibfnamefont {W.}~\bibnamefont {Zhang}},\ }\href
  {http://dx.doi.org/10.1186/s11671-015-0914-3} {\bibfield  {journal} {\bibinfo
   {journal} {Nanoscale Res. Lett.}\ }\textbf {\bibinfo {volume} {10}},\
  \bibinfo {pages} {223} (\bibinfo {year} {2015})}\BibitemShut {NoStop}%
\bibitem [{\citenamefont {Bara\'nski}\ \emph {et~al.}(2017)\citenamefont
  {Bara\'nski}, \citenamefont {Kobia{\l}ka},\ and\ \citenamefont
  {Doma\'nski}}]{Baranski2017}%
  \BibitemOpen
  \bibfield  {author} {\bibinfo {author} {\bibfnamefont {J.}~\bibnamefont
  {Bara\'nski}}, \bibinfo {author} {\bibfnamefont {A.}~\bibnamefont
  {Kobia{\l}ka}}, \ and\ \bibinfo {author} {\bibfnamefont {T.}~\bibnamefont
  {Doma\'nski}},\ }\href {http://stacks.iop.org/0953-8984/29/i=7/a=075603}
  {\bibfield  {journal} {\bibinfo  {journal} {J. Phys. Condens. Matter}\ }\textbf {\bibinfo {volume} {29}},\ \bibinfo {pages} {075603}
  (\bibinfo {year} {2017})}\BibitemShut {NoStop}%
\bibitem [{\citenamefont {Xiong}(2016)}]{Xiong2016}%
  \BibitemOpen
  \bibfield  {author} {\bibinfo {author} {\bibfnamefont {Y.}~\bibnamefont
  {Xiong}},\ }\href {\doibase 10.1088/0256-307x/33/5/057402} {\bibfield
  {journal} {\bibinfo  {journal} {Chinese Phys. Lett.}\ }\textbf {\bibinfo
  {volume} {33}},\ \bibinfo {pages} {057402} (\bibinfo {year}
  {2016})}\BibitemShut {NoStop}%
\bibitem [{\citenamefont {Ueda}\ and\ \citenamefont
  {Yokoyama}(2014)}]{Ueda2014}%
  \BibitemOpen
  \bibfield  {author} {\bibinfo {author} {\bibfnamefont {A.}~\bibnamefont
  {Ueda}}\ and\ \bibinfo {author} {\bibfnamefont {T.}~\bibnamefont
  {Yokoyama}},\ }\href {http://dx.doi.org/10.1103/PhysRevB.90.081405}
  {\bibfield  {journal} {\bibinfo  {journal} {Phys. Rev. B}\ }\textbf {\bibinfo
  {volume} {90}},\ \bibinfo {pages} {081405} (\bibinfo {year}
  {2014})}\BibitemShut {NoStop}%
\bibitem [{\citenamefont {Jiang}\ and\ \citenamefont
  {Zheng}(2015)}]{Jiang2015}%
  \BibitemOpen
  \bibfield  {author} {\bibinfo {author} {\bibfnamefont {C.}~\bibnamefont
  {Jiang}}\ and\ \bibinfo {author} {\bibfnamefont {Y.-S.}\ \bibnamefont
  {Zheng}},\ }\href {\doibase 10.1016/j.ssc.2015.04.001} {\bibfield  {journal}
  {\bibinfo  {journal} {Solid State Commun.}\ }\textbf {\bibinfo {volume}
  {212}},\ \bibinfo {pages} {14–18} (\bibinfo {year} {2015})}\BibitemShut
  {NoStop}%
\bibitem [{\citenamefont {Zeng}\ \emph {et~al.}()\citenamefont {Zeng},
  \citenamefont {Chen}, \citenamefont {You},\ and\ \citenamefont
  {Lv}}]{Zeng2015}%
  \BibitemOpen
  \bibfield  {author} {\bibinfo {author} {\bibfnamefont {Q.-B.}\ \bibnamefont
  {Zeng}}, \bibinfo {author} {\bibfnamefont {S.}~\bibnamefont {Chen}}, \bibinfo
  {author} {\bibfnamefont {L.}~\bibnamefont {You}}, \ and\ \bibinfo {author}
  {\bibfnamefont {R.}~\bibnamefont {L\"u}},\ }\href@noop {} {\ }\Eprint
  {http://arxiv.org/abs/arXiv:1510.06558} {arXiv:1510.06558} \BibitemShut
  {NoStop}%
\bibitem [{\citenamefont {Levitov}\ and\ \citenamefont
  {Reznikov}(2004)}]{Levitov2004}%
  \BibitemOpen
  \bibfield  {author} {\bibinfo {author} {\bibfnamefont {L.~S.}\ \bibnamefont
  {Levitov}}\ and\ \bibinfo {author} {\bibfnamefont {M.}~\bibnamefont
  {Reznikov}},\ }\href {http://dx.doi.org/10.1103/PhysRevB.70.115305}
  {\bibfield  {journal} {\bibinfo  {journal} {Phys. Rev. B}\ }\textbf {\bibinfo
  {volume} {70}},\ \bibinfo {pages} {115305} (\bibinfo {year}
  {2004})}\BibitemShut {NoStop}%
\bibitem [{\citenamefont {Weithofer}\ \emph {et~al.}(2014)\citenamefont
  {Weithofer}, \citenamefont {Recher},\ and\ \citenamefont
  {Schmidt}}]{Weithofer2014}%
  \BibitemOpen
  \bibfield  {author} {\bibinfo {author} {\bibfnamefont {L.}~\bibnamefont
  {Weithofer}}, \bibinfo {author} {\bibfnamefont {P.}~\bibnamefont {Recher}}, \
  and\ \bibinfo {author} {\bibfnamefont {T.~L.}\ \bibnamefont {Schmidt}},\
  }\href {\doibase 10.1103/physrevb.90.205416} {\bibfield  {journal} {\bibinfo
  {journal} {Phys. Rev. B}\ }\textbf {\bibinfo {volume} {90}},\ \bibinfo
  {pages} {205416} (\bibinfo {year} {2014})}\BibitemShut {NoStop}%
\bibitem [{\citenamefont {Fano}(1935)}]{Fano1935}%
  \BibitemOpen
  \bibfield  {author} {\bibinfo {author} {\bibfnamefont {U.}~\bibnamefont
  {Fano}},\ }\href@noop {} {\bibfield  {journal} {\bibinfo  {journal} {Nuovo
  Cimento}\ }\textbf {\bibinfo {volume} {12}},\ \bibinfo {pages} {154}
  (\bibinfo {year} {1935})}\BibitemShut {NoStop}%
\bibitem [{\citenamefont {Clarke}()}]{Clarke2017}%
  \BibitemOpen
  \bibfield  {author} {\bibinfo {author} {\bibfnamefont {D.~J.}\ \bibnamefont
  {Clarke}},\ }\href@noop {} {\bibinfo  {journal} {arXiv:1702.01740}\
  }\BibitemShut {NoStop}%
\bibitem [{\citenamefont {Prada}\ \emph {et~al.}()\citenamefont {Prada},
  \citenamefont {Aguado},\ and\ \citenamefont {San-Jos\'{e}}}]{Prada2017}%
  \BibitemOpen
\bibfield  {journal} {  }\bibfield  {author} {\bibinfo {author} {\bibfnamefont
  {E.}~\bibnamefont {Prada}}, \bibinfo {author} {\bibfnamefont
  {R.}~\bibnamefont {Aguado}}, \ and\ \bibinfo {author} {\bibfnamefont
  {P.}~\bibnamefont {San-Jos\'{e}}},\ }\href@noop {} {\bibinfo  {journal}
  {arXiv:1702.02525}\ }\BibitemShut {NoStop}%
\bibitem [{\citenamefont {Zazunov}\ \emph {et~al.}(2016)\citenamefont
  {Zazunov}, \citenamefont {Egger},\ and\ \citenamefont
  {Levy~Yeyati}}]{PhysRevB.94.014502}%
  \BibitemOpen
\bibfield  {journal} {  }\bibfield  {author} {\bibinfo {author} {\bibfnamefont
  {A.}~\bibnamefont {Zazunov}}, \bibinfo {author} {\bibfnamefont
  {R.}~\bibnamefont {Egger}}, \ and\ \bibinfo {author} {\bibfnamefont
  {A.}~\bibnamefont {Levy~Yeyati}},\ }\href {\doibase
  10.1103/PhysRevB.94.014502} {\bibfield  {journal} {\bibinfo  {journal} {Phys.
  Rev. B}\ }\textbf {\bibinfo {volume} {94}},\ \bibinfo {pages} {014502}
  (\bibinfo {year} {2016})}\BibitemShut {NoStop}%
\bibitem [{\citenamefont {Pikulin}\ and\ \citenamefont
  {Nazarov}(2012)}]{Pikulin2012}%
  \BibitemOpen
  \bibfield  {author} {\bibinfo {author} {\bibfnamefont {D.~I.}\ \bibnamefont
  {Pikulin}}\ and\ \bibinfo {author} {\bibfnamefont {Y.~V.}\ \bibnamefont
  {Nazarov}},\ }\href {\doibase 10.1134/S0021364011210090} {\bibfield
  {journal} {\bibinfo  {journal} {JETP Letters}\ }\textbf {\bibinfo {volume}
  {94}},\ \bibinfo {pages} {693} (\bibinfo {year} {2012})}\BibitemShut
  {NoStop}%
\end{thebibliography}
\end{document}